\documentclass[11pt, a4paper,onecolumn]{belarticle4}
\pdfoutput=1
\usepackage{revsymb, amsmath, amsfonts, amssymb, enumerate, fullpage, amsthm, graphicx, braket, relsize, bbm, mathrsfs, mathtools}
\usepackage{graphicx,color}
\definecolor{darkbrown}{rgb}{0.4, 0.26, 0.13}
\definecolor{ao}{rgb}{0.0, 0.5, 0.0}
\definecolor{bleudefrance}{rgb}{0.19, 0.55, 0.91}
\usepackage{adjustbox}

\usepackage{paralist}
\usepackage{subfigure}

\usepackage{tikz}
\usetikzlibrary{calc,decorations.pathreplacing}
\usetikzlibrary{arrows,shapes}

\usepackage[linktocpage=true, colorlinks=true, linkcolor=blue, urlcolor=blue, citecolor=blue]{hyperref}

\newtheorem{theo}{Theorem}

\newtheorem{prop}[theo]{Proposition} 
\newtheorem{lem}[theo]{Lemma}
\newtheorem{cor}[theo]{Corollary}

\newtheorem{defn}[theo]{Definition}

\newcommand{\cH}{\mathcal{H}}

\newcommand{\id}{\mathbb{I}}
\newcommand{\tr}[2]{\mathrm{tr}_{#2} \left\{ #1 \right\}}

\newcommand{\MN}{\mathbf{MutNorm}}
\newcommand{\AOL}{\mathbf{AltOrLab}}
\newcommand{\TB}{\mathrm{TH}}

\usepackage{cite}

\usepackage{tcolorbox}
\tcbuselibrary{skins,breakable}
\usetikzlibrary{shadings,shadows}

    {\endtcolorbox}

\newcommand{\overbar}[1]{\mkern 1.5mu\overline{\mkern-1.5mu#1\mkern-1.5mu}\mkern 1.5mu}
\newcommand{\disj}{\mathrm{disj}}
\newcommand{\comm}{\mathrm{comm}}

\definecolor{darkgreen}{rgb}{0,.5,0}

\definecolor{orangy}{RGB}{213,94,0}

\begin{document}

\title{On almost-quantum correlations and the composition of contextuality scenarios}


\author{Ana Belén Sainz}
\affiliation{International Centre for Theory of Quantum Technologies, University of  Gda{\'n}sk, 80-309 Gda{\'n}sk, Poland} 
\affiliation{Basic Research Community for Physics e.V., Germany}


\begin{abstract}
Within the hypergraph framework for contextuality, probability normalisation and orthogonality of events play a central role in characterising physical behaviours and composing operational scenarios. Here, we revisit the interplay between these two notions and disprove three conjectures concerning the structural properties of contextuality scenarios and their composition. We show that equality of orthogonality graphs is neither necessary nor sufficient for two scenarios to admit the same set of almost-quantum behaviours. We establish necessary and sufficient conditions for two contextuality scenarios to have the same sets of quantum behaviours or the same sets of almost-quantum behaviours. With this we articulate a qualitative difference between these two types of behaviours, that may inform the search for physical principles distinguishing them. Regarding composition, we show that neither the Foulis–Randall product nor its disjunctive product variant preserve self-completeness in general, and provide an efficiently-verifiable sufficient condition for the latter to do so. Finally, we discuss how considering information about the relations among local measurements -- and hence leaving the Bell paradigm -- impacts the recently established correspondence between hypergraph products and process functions. Indeed, already in bipartite settings, quasi-process functions may emerge from the additional composite hyperedges already within the disjunctive FR-product.
\end{abstract}

\maketitle

\tableofcontents

\section{Introduction}\label{sec:intro}

It is now a widely accepted fact that nature defies a classical explanation and that quantum theory explains the observed phenomenology remarkably well. The scope of quantum theory, however, is still not fully understood -- a question that impacts the development of technologies that harness quantum advantage. Research effort has been invested in the last century to reframe relevant phenomenology of quantum theory in terms of intuitive physical or information-theoretic principles, and move away from its abstract mathematical axiomatisation. One particular case is that of Bell \cite{bell64} or Kochen-Specker \cite{ks} scenarios:  non-classical correlations in Bell scenarios are crucial for some quantum technologies \cite{brunner2014bell}, and can be understood as a special case of the broader class of experiments formalised by Kochen and Specker \cite{acin2015combinatorial}. Several principles have been put forward to characterise quantum correlations and single them out from the landscape of possibilities accommodated by special relativity \cite{van2005implausible,brassard2006limit,pawlowski2009information,linden2007quantum,navascues2010glance,fritz2013local,acin2015combinatorial}. These attempts have not been successful however, due to the existence of a mathematically-defined set of correlations called almost-quantum: these don't violate any of the principles proposed so far, and they strictly contain the set of quantum correlations in some scenarios such as Bell's \cite{navascues2015almost}. 

The hypergraph framework of Ref.~\cite{acin2015combinatorial} provides a common language in which to jointly study Bell and Kochen-Specker scenarios. This formalism focuses particularly on how to compose such contextuality scenarios in a meaningful way, so that physical considerations are structurally encoded in the corresponding hypergraphs. 
Two properties of contextuality scenarios play a crucial role in this framework: the normalisation of probabilities associated with measurement outcomes, and the orthogonality relations among events.
Different composition rules have been explored based on these properties \cite{acin2015combinatorial,sainz2018multipartite}, and have had an impact beyond the study of contextuality itself. Indeed, recent work in the field of causality has shown how hyperedges that arise within a particular composition of $n$ Bell scenarios are in one-to-one correspondence with the valid process functions on $n$ devices \cite{leitherer2026bell}. 

In this work we explore the relation between the normalisation and orthogonality conditions in the hypergraph framework of Ref.~\cite{acin2015combinatorial}, and how this may shed light both on the nature of nonclassical correlations and on the operational meaning of multipartite composition. We focus on how the normalisation and orthogonality conditions of one scenario relate to those of another, and study their role in enabling the two scenarios to have the same set of quantum correlations or of almost-quantum ones. Through this, we find a qualitative difference in the role that these notions play in characterising quantum and almost-quantum behaviours. We also provide necessary and sufficient conditions for two scenarios to have the same set of quantum or almost-quantum predictions, and explore numerical tools to certify this. In doing so, we disprove a conjecture put forward in Ref.~\cite{sainz2018multipartite} on the role that orthogonality plays for rendering almost-quantum sets equal. 

We moreover explore how the connection between composition rules and process functions behaves when the hypergraphs being composed are not Bell scenarios. Bell scenarios, as typically considered in the device-independent paradigm, do not incorporate information about the operational relations among outcomes of different local measurements. This information -- which can be naturally accommodated within the hypergraph framework -- can be exploited to strengthen  nonclassicality certification and characterisation \cite{porto2024trade,mazzari2023generalized,xue2023synchronous,temistocles2019measurement}. In this paper we show how this extra information implies that the disjunctive composition rule of general contextuality scenarios includes hyperedges that correspond to quasi-process functions which feature both the grandfather antinomy and the bootstrap antinomy. In doing so, we disprove two conjectures put forward in Ref.~\cite{sainz2018multipartite} on when and how the `completed composition' can be computed from the disjunctive composition. Finally, we provide a sufficient condition for the disjunctive Foulis–Randall product to yield a self-complete scenario. 

Our results motivate further exploration of the physical principles underlying almost-quantum phenomena and of the interplay between local contextuality and multipartite causal structures.

\section{Preliminaries}\label{sec:prel}

\subsection{The hypergraph framework and multipartite composition of contextuality scenarios}

In Bell and Kochen-Specker experiments, one has a physical system on which measurements are performed and outcomes are obtained. The object of study is then the conditional probability distribution on these outcomes. What Bell and Kochen-Specker scenarios differ on is the operational constraints that the outcome probabilities must satisfy. On the one hand, for Bell, the outcomes correspond to collections of outcomes of local measurements performed in a space-like manner by distant parties, and hence the constraint the probabilities must satisfy pertain to the No Signalling principle. On the other hand, for Kochen-Specker, the constraints are more subtle, and pertain to the compatibility of the measurements being performed on the system, or on the operational equivalences featured among outcomes of different measurements. 
Ref.~\cite{acin2015combinatorial} develops a framework to represent operational scenarios pertaining to Bell and Kochen-Specker experiments, in a way that these operational constraints are fully captured by structural features of the hypergraphs through which these scenarios are represented. Our aim here is not to review how to mathematically formalise these physical situations, so we refer the reader to Refs.~\cite{acin2015combinatorial,fritz2013probabilistic} for more details. 

Bell and Kochen-Specker operational scenarios are then formalised via a hypergraph $H = (V,E)$. Its  vertices $V$ (also referred to as \textit{events}) correspond to the collection of all the outcomes of all the (finite number of) measurements we may perform on the systems (specified in each experiment). Each such measurement then is represented in $H$ via a hyperedge in $E$, which contains all the possible outcomes of the measurement it corresponds to (conversely, each hyperedge in $E$ must represent one such measurement). The object of study in these scenarios is the conditional probability distributions $\mathbf{p}_H := \{p(v)\}_{v \in V}$, which we also denote as \textit{behaviours}. For such a behaviour to be admissible in any physical theory, it must satisfy the linear constraints $\sum_{v\in e} p(v) =1 \quad \forall \, e \in E$, which stem from the completeness of the measurements being performed. The set of all possible behaviours for a given scenario $H$ that satisfy these normalisation constraints is denoted by $\mathcal{G}(H)$. Another useful concept is that of an \textit{orthogonality graph}\footnote{This is called \textit{adjacency graph} in the formalism of Ref.~\cite{cabello2010non}.} associated to a contextuality scenario $H$, which we denote by $\mathrm{OrthoGraph}(H)$ \cite{fritz2013local,sainz2014exploring}. This graph has the same vertex set as $H$, and its edges are defined as follows: $\{u,v\} \in E(\mathrm{OrthoGraph}(H))$ iff $\exists \, e \in E(H) \,$ s.t.~$ \,u,v\in e$\,. In addition, one can define the \textit{non-orthogonality graph} of a scenario $H$ \cite{acin2015combinatorial}, denoted by $\mathrm{NO}(H)$, as the graph complement of $\mathrm{OrthoGraph}(H)$.

\begin{figure}
\begin{center}
\subfigure[Single-party Bell scenario $\mathcal{B}_{1,2,2}$]{
\begin{tikzpicture}[scale=1]

	\node[draw,shape=circle,fill,scale=.5] (e) at (0,0.2) {} ;
	\node[above=0pt] at (e.north) {\tiny{$0|0$}} ;
	\node[draw,shape=circle,fill,scale=.5] (f) at (2.5,0.2) {} ;
	\node[above=0pt] at (f.north) {\tiny{$1|0$}} ;
	\draw[thick,darkgreen] (1.25,0.2) ellipse (2cm and .3cm) ;
	
	\node[draw,shape=circle,fill,scale=.5] (e) at (0,-1.9) {} ;
	\node[above=0pt] at (e.north) {\tiny{$0|1$}} ;
	\node[draw,shape=circle,fill,scale=.5] (f) at (2.5,-1.9) {} ;
	\node[above=0pt] at (f.north) {\tiny{$1|1$}} ;
	\draw[thick,darkgreen] (1.25,-1.9) ellipse (2cm and .3cm) ;
    \node[draw, color=white] at (-1,-2.5) {};
    \node[draw, color=white] at (4,0) {};

\end{tikzpicture}
} \hskip 1.5cm
\subfigure[{Bipartite Bell scenario  $\mathcal{B}_{2,2,2} = \mathcal{B}_{1,2,2} {}^{\mathrm{FR}}\otimes \mathcal{B}_{1,2,2}$}]{
		\begin{tikzpicture}[scale=1]
		\clip (0,.5) rectangle (5,5);
		\foreach \x in {0,1} \foreach \y in {0,1} \foreach \a in {0,1} \foreach \b in {0,1}
		{
			\node[draw,shape=circle,fill,scale=.5] (e) at (\b+2*\y+1,4-\a-2*\x) { } ;
			\node[above=0pt] at (e) {\tiny{$\a\b|\x\y$}} ;
		}
		\foreach \y in {2} \draw[thick,red,rounded corners] (2.8,\y+0.8) -- (4.33,\y+0.8) -- (4.33,\y+1.2) -- (2.8,\y+1.2) -- (2.25,\y+2.2) -- (0.67,\y+2.2) -- (0.67,\y+1.8) -- (2.2,\y+1.8) -- cycle ;		
				
		\foreach \x in {0,2} \foreach \y in {0,2} \draw[rounded corners,thick,blue] (\x+0.55,\y+0.68) rectangle (\x+2.45,\y+2.52) ;
		\foreach \y in {0,2} \draw[thick,red,rounded corners] (2.8,\y+0.8) -- (4.33,\y+0.8) -- (4.33,\y+1.2) -- (2.8,\y+1.2) -- (2.25,\y+2.2) -- (0.67,\y+2.2) -- (0.67,\y+1.8) -- (2.2,\y+1.8) -- cycle ;
		\foreach \y in {0,2} \draw[thick,red,rounded corners] (2.8,\y+1.8) -- (4.33,\y+1.8) -- (4.33,\y+2.2) -- (2.75,\y+2.2) -- (2.2,\y+1.2) -- (0.67,\y+1.2) -- (0.67,\y+0.8) -- (2.2,\y+0.8) -- cycle ;
		\foreach \x in {0,2} \draw[thick,darkgreen,rounded corners] (\x+0.75,2.2) -- (\x+0.75,0.73) -- (\x+1.25,0.73) -- (\x+1.25,2.2) -- (\x+2.25,2.9) -- (\x+2.25,4.2) -- (\x+1.78,4.2) -- (\x+1.78,2.9) -- cycle ;
		\foreach \x in {0,2} \draw[thick,darkgreen,rounded corners] (5-\x-0.75,2.2) -- (5-\x-0.75,0.73) -- (5-\x-1.25,0.73) -- (5-\x-1.25,2.2) -- (5-\x-2.25,2.9) -- (5-\x-2.25,4.2) -- (5-\x-1.78,4.2) -- (5-\x-1.78,2.9) -- cycle ;
        \node[draw, color=white] at (-50,0) {};
        \node[draw, color=white] at (100,0) {};
		\end{tikzpicture} 
}    
\end{center}
\caption{Example of a bipartite Bell scenario represented as a composition of hypergraphs. (a) Each party has two dichotomic measurements, which is depicted by the hypergraph $\mathcal{B}_{1,2,2}$ that has no overlapping hyperedges. (b) The hypergraph representing the composite bipartite scenario is given by $\mathcal{B}_{2,2,2}$ as the Foulis-Randall product of $\mathcal{B}_{1,2,2}$ with itself.  }
\label{fig:bell}
\end{figure}
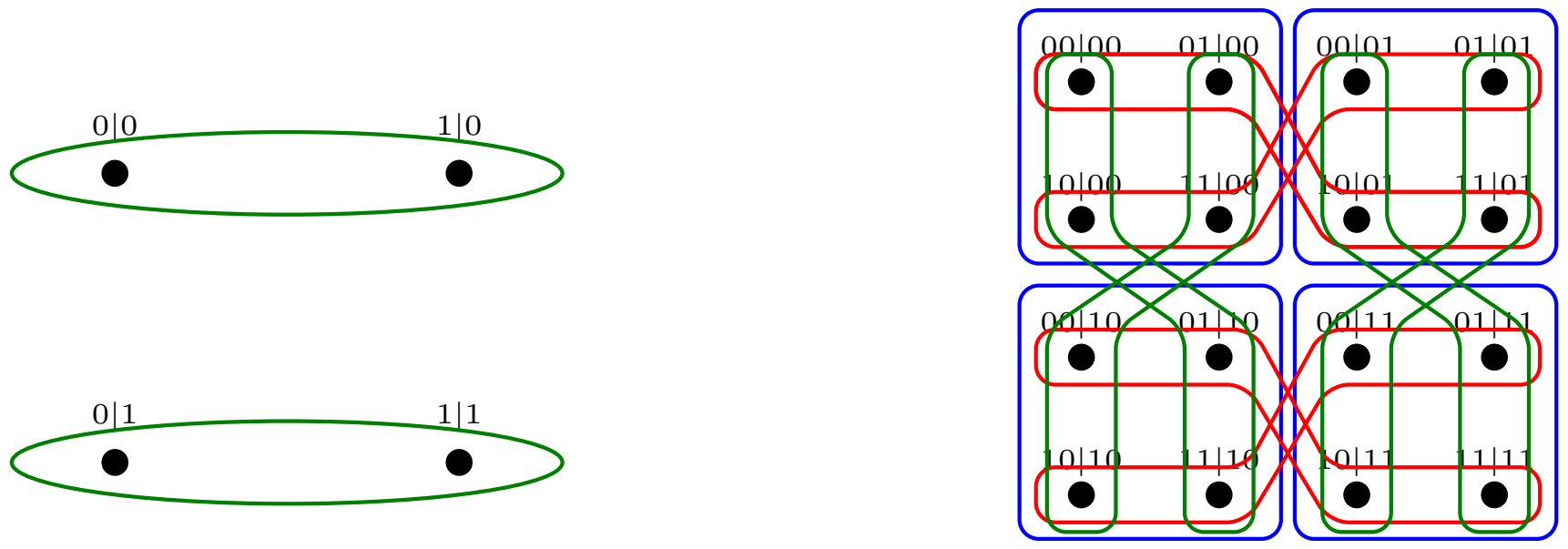

One natural question is, given two independent scenarios $H_A$ and $H_B$, how can these be represented jointly in a single hypergraph $H_{AB}$ in a way that the linear constraints imposed on the joint scenario meaningfully capture the physics underpinning them (such as space-like separation, see Ref.~\cite{acin2015combinatorial}). In order to answer this question, many composition rules -- in the form of hypergraph products -- have been proposed. These include: 
\begin{compactitem}
\item the \textit{minimal} FR-product, denoted  $\quad^{\min}\otimes$,
\item the \textit{common} FR-product, denoted $\quad^{\comm}\otimes$,
\item the \textit{maximal} FR-product, denoted  $\quad^{\max}\otimes$,
\item the \textit{disjunctive} FR-product, denoted $\quad^{\disj}\otimes$,
\item and the \textit{completed} FR-product, denoted $\quad\overbar{\otimes}$,
\end{compactitem}
where we have used the notation of Ref.~\cite{sainz2018multipartite}. For their definitions we refer the reader to Ref.~\cite{sainz2018multipartite}.
In all these composition rules, the set of vertices for the composite coincide, and is given by the cartesian product of the vertex sets of the individual hypergraphs. What these composition rules differ in is what subsets of the composite vertex set correspond to formal hyperedges. As an example, Fig.~\ref{fig:bell} presents the hypergraph that represents a bipartite Bell scenario, where each party has access to two dichotomic local measurements: $\mathcal{B}_{1,2,2}$ is the hypergraph for each party (Alice and Bob), and  $\mathcal{B}_{2,2,2} = \mathcal{B}_{1,2,2} \, {}^{\comm}\otimes \mathcal{B}_{1,2,2}$ is a hypergraph whose hyperedges capture all nonsignalling constraints. Here we use the notation $\mathcal{B}_{1,m,d}$ to denote a local Bell scenario consisting of a single party with $m$ measurement choices of $d$ outcomes each, which is represented as a hypergraph with $m$ disjoint hyperedges each containing $d$ vertices.

Ref.~\cite{sainz2018multipartite} put forward the following conjecture, which was proven to hold for the special case of Bell scenarios \cite{leitherer2026bell}, i.e., when each individual scenario $H_k$ is of the form $H_k=\mathcal{B}_{1,m_k,d_k}$: 

\bigskip

\noindent \textbf{Conjecture [\# 2 of Ref.~\cite{sainz2018multipartite}]}: $\overbar{\otimes}_{i=1}^n \overbar{H}_i =  {}^{\disj}\otimes _{i=1}^n \overbar{H}_i$. In the case that the individual hypergraphs being composed are themselves self-complete (i.e., $\overbar{H}_i = H_i$), this translates to the disjunctive FR-product giving rise to a self-complete composite hypergraph. 

\bigskip

Another conjecture in Ref.~\cite{sainz2018multipartite} states the following: 

\bigskip

\noindent \textbf{Conjecture [Footnote \# 10 of Ref.~\cite{sainz2018multipartite}]}: If 
$\overbar{H}_A = H_A$ and $\overbar{H}_B = H_B$, then $H_A \overbar{\otimes} H_B = H_A {}^{\mathrm{FR}}\otimes H_B$. Here ${}^{\mathrm{FR}}\otimes$ denotes the standard FR-product, which for bipartite scenarios coincides with these three products: ${}^{\min}\otimes$, ${}^{\comm}\otimes$, and ${}^{\max}\otimes$.

\bigskip

In Sec.~\ref{se:counter} we will show that these two  conjectures are not valid for arbitrary contextuality scenarios. 

\subsection{Quantum and almost-quantum behaviours}

Almost-quantum behaviours were first studied in the context of Bell scenarios \cite{navascues2015almost}, and later generalised to arbitrary Kochen-Specker scenarios \cite{acin2015combinatorial}. Their physical relevance stems from the fact that they do not violate any of the principles defined so far to characterise quantum behaviours, and hence pose the main challenge to that research program. In turn, almost-quantum behaviours are easily characterised in terms of a semidefinite program, and have become a versatile computational tool for research in foundations and quantum information (e.g., to certify maximum quantum violation or post-quantumness of operational phenomena) \cite{padovan2024secure,liu2024almost,adamson2020quantum,tavakoli2023semidefinite}. 

Given a contextuality scenario $H$, the set of almost-quantum behaviours that it admits is characterised in different equivalent ways (see Ref.~\cite[Prop.~6.3.1]{acin2015combinatorial} and Ref.~\cite[Prop.~6.3.2]{acin2015combinatorial}). A particular characterisation that we will leverage in this manuscript is the following: 


\begin{defn}\label{def:aqaflsc}
(Ref.~\cite[Prop.~6.3.1(c)]{acin2015combinatorial}) \\
Given a contextuality scenario $H$, a behaviour $\mathbf{p} \in \mathcal{G}(H)$ admits an almost-quantum model, iff
there exists a Hilbert space $\mathcal{H}$, a unit vector $\ket{\Psi} \in \mathcal{H}$, and unit vectors $\ket{v}$ for each $v \in V(H)$ such that: 
\begin{align}
    &p(v) = \tr{P_v \, \ket{\Psi}\bra{\Psi}}{} \quad  \forall  \, v \in V(H)\,,\\
    &u \perp_H v \, \Rightarrow \, \langle u \vert v \rangle = 0 \quad \forall \, u,v \in V(H)\,,
\end{align}
where $P_v = \ket{v}\bra{v}$ and $\perp_H$ denotes orthogonality of the two vertices $u,v$; i.e., there exists a hyperedge in $H$ where they both belong to. 

The set of all almost-quantum behaviours admitted by a scenario $H$ is denoted by $\mathcal{Q}_1(H)$.
\end{defn}

\bigskip

Of particular relevance to this work is a conjecture from Ref.~\cite{sainz2018multipartite}, which we disprove in Sec.~\ref{se:motex}:

\bigskip

\noindent \textbf{Conjecture [\# 1 of Ref.~\cite{sainz2018multipartite}]}: 
If $\mathrm{OrthoGraph}(H_1) \neq \mathrm{OrthoGraph}(H_2)$ then $\mathcal{Q}_1(H_1) \neq \mathcal{Q}_1(H_2)$, even if $\overbar{H}_1 = \overbar{H}_2$.

\bigskip

The set of quantum behaviours -- those that quantum theory allows via projective measurements on physical systems -- is included within the set of almost-quantum behaviours (sometimes strictly, like in Bell scenarios \cite{navascues2015almost}), and its mathematical definition is as follows:

\begin{defn}\label{def:qafls}
(Ref.~\cite[Def.~5.1.1]{acin2015combinatorial}) \\
Given a contextuality scenario $H$, a behaviour $\mathbf{p} \in \mathcal{G}(H)$ admits a quantum model, iff
there exists a Hilbert space $\mathcal{H}$, a density matrix $\rho$ in $\mathcal{H}$, and a projector $P_v$ for each $v \in V(H)$ such that: 
\begin{align}
    &p(v) = \tr{P_v \, \rho}{} \quad \forall \, v \in V(H)\,,\\
    & \sum_{v \in e} P_v = \id\quad \forall \, e \in E(H)\,. \label{eq:qnorP}
\end{align}
The set of all quantum behaviours admitted by a scenario $H$ is denoted by $\mathcal{Q}(H)$. 
\end{defn}

\bigskip

Throughout the manuscript, we will refer to these states, vectors, and/or projectors as the ones providing a \textit{model} for the quantum or almost-quantum behaviour being considered.

\subsection{Causality and hypergraph products}

In this section we will briefly summarise the results of Ref.~\cite{acin2015combinatorial} and Ref.~\cite{leitherer2026bell} relevant for this paper. These pertain to the operational and physical interpretation of the hyperedges that arise via the different composition rules mentioned in the previous subsection. 

As discussed in Refs.~\cite{acin2015combinatorial, sainz2018multipartite}, the minimal, common, and maximal FR-products have interpretations in terms of \textit{measurement protocols} (a.k.a.~correlated measurements)  among the parties, where by the word \textit{party} we denote any single contextuality scenario being composed. A measurement protocol in a composite scenario (let us focus on the case of a bipartite composite for simplicity) consists of choosing a hyperedge from the hypergraph of one of the parties (i.e., a party choosing a measurement) and then associating to each of its outcomes a hyperedge from the hypergraph of the other party. In the case of Bell scenarios, this corresponds to Alice making a measurement, obtaining an outcome, and then Bob choosing what measurement to perform as a function of Alice's outcome. The above-mentioned composition rules then have the following interpretation: 
\begin{compactitem}
\item the minimal FR-product involves all-and-only hyperedges corresponding to measurement protocols of only two `time steps': in the first step, all-but-one of the parties chooses a measurement (hyperedge), and in the second step the remaining party chooses which measurement (hyperedge) to perform depending on the outcome of the other parties in the first step. Generating all hyperedges of ${}^{\min}\otimes H_k$ then involves going through all possible choices of parties for the first time step, all choices of hyperedges for those parties, and all choices of functions for the last party (through which she decides which hyperedge to measure according to the collective output of the previous parties). 
\item the common FR-product involves all-and-only hyperedges corresponding to measurement protocols of $n$ `time steps' -- where $n$ is the number of hypergraphs being composed -- and where the order in which the parties measure is fixed: at each time step, the party that is due to act then receives the outcomes obtained by all the previous parties in the protocol, uses them to choose a measurement (hyperedge), implements it, obtains an outcome, and communicates it (together with the other outcomes) to the next party in the protocol. Generating all hyperedges of ${}^{\comm}\otimes H_k$ then involves, among other things, going through all possible fixed orderings of the parties. 
\item the maximal FR-product involves all-and-only hyperedges corresponding to measurement protocols of $n$ `time steps' and where the order in which the parties measure is decided dynamically depending on the outcomes obtained: at each time step, the party that is due to act then receives the outcomes obtained by all the previous parties in the protocol, uses them to choose a measurement (hyperedge), implements it, obtains an outcome, decides which party goes next in the protocol, and  communicates to that party all the measurement outcomes obtained so far (hers included). Generating all hyperedges of ${}^{\max}\otimes H_k$ then involves, among other things, going through all possible ways to dynamically choose the order of the parties. 
\end{compactitem}

\bigskip

Ref.~\cite{leitherer2026bell} provided an interpretation of the disjunctive FR-product of local Bell scenarios in terms of process functions \cite{baumeler2016space,baumeler2016device,baumeler2019reversible}, as well as a reinterpretation of the minimal, common, and maximal FR-products of Bell scenarios in that language. Process functions can be understood as classical deterministic counterparts of process matrices \cite{oreshkov2012quantum}, that is, as the most general multipartite classical protocol  where `locally a party performs a measurement on her system and obtains an outcome' and no time-travel paradoxes are generated. Process functions then allow for processes where the order in which the parties act may be undetermined (i.e., something more general than the measurement protocols given by the maximal FR-product), as illustrated by the Lugano process \cite{baumeler2016device}. In this language, Ref.~\cite{leitherer2026bell} shows the following: 
\begin{compactitem}
\item the hyperedges of the minimal FR-product of local Bell scenarios correspond to `all-to-one communication' process functions. In turn, they correspond to the deterministic and classical version of a quantum circuit with operations used in parallel (QC-PAR). 
\item the hyperedges of the common FR-product of local Bell scenarios correspond to `fixed causal order' process functions. In turn, they correspond to the deterministic and classical version of a quantum circuit with fixed causal order (QC-FO), known as quantum combs. 
\item the hyperedges of the  maximal FR-product of local Bell scenarios  correspond to causal process functions. In turn, they correspond to the deterministic and classical version of a quantum circuit with classical control of causal order (QC-CC). 
\item the hyperedges of the disjunctive FR-product of local Bell scenarios correspond to all valid process functions, and have no counterpart in the quantum circuit hierarchy. 
\end{compactitem}
Ref.~\cite{leitherer2026bell} shows that valid process functions are isomorphic to the hyperedges of the disjunctive FR-product of local Bell scenarios $\mathcal{B}_{1,m_k,d_k}$, and their proof technique relies on proving that Conjecture 2 of Ref.~\cite{sainz2018multipartite} is true for the composition of local Bell scenarios.
Open questions include whether Conjecture 2 of Ref.~\cite{sainz2018multipartite} holds in general, and whether the hyperedges of the disjunctive FR-product of more general scenarios include quasi-process functions (i.e., non-valid ones).

\section{Orthogonality graphs and almost-quantum behaviours}\label{se:motex}

In this section we are interested in whether the sets of almost-quantum behaviours  coincide for different contextuality scenarios, and how this may or not depend on the way the two scenarios relate to each other. We present a series of examples which showcase the subtle role that orthogonality relations play in the emergent almost-quantum behaviours. 

\bigskip

We begin by highlighting that two contextuality scenarios having the same orthogonality graph do not necessarily admit the same set of almost-quantum behaviours, especially when their hypergraphs have different completions. 

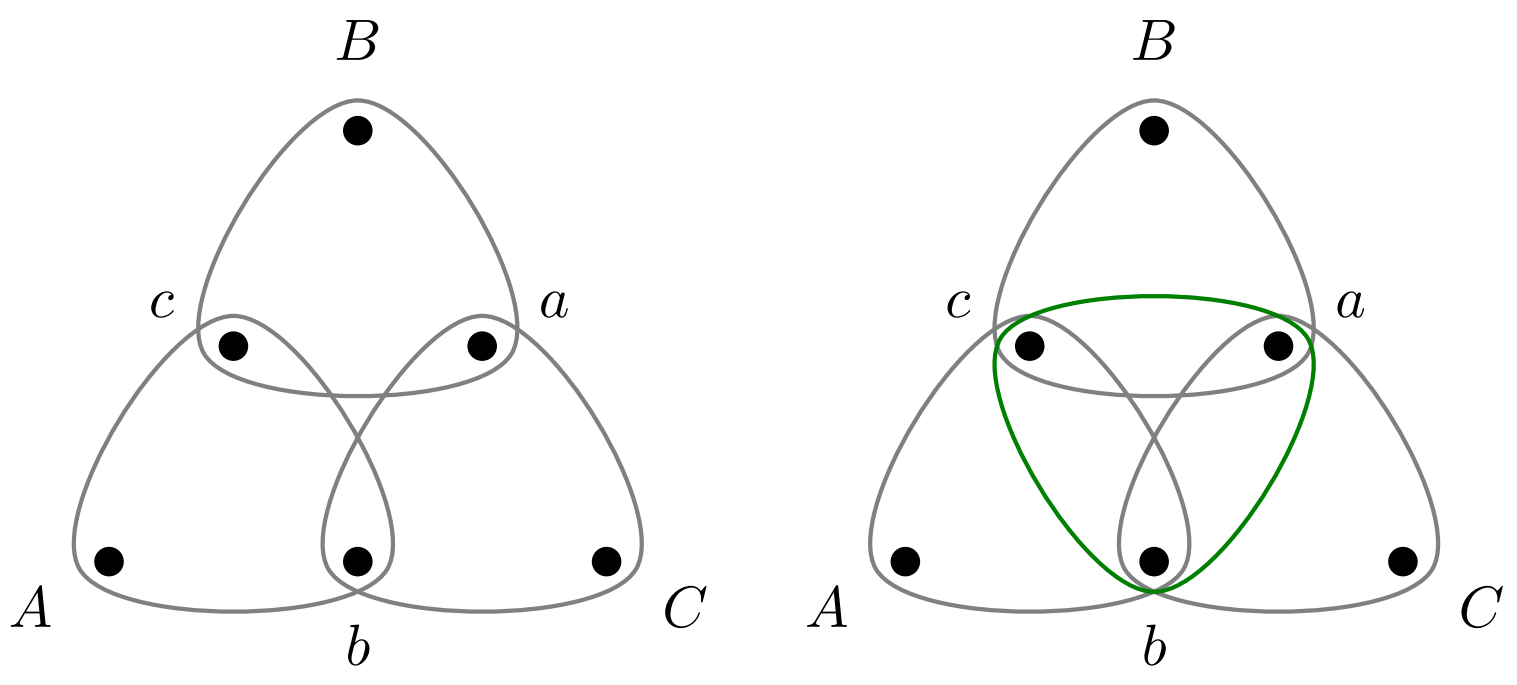
\begin{figure}
\centering
\subfigure[Scenario $H_1$]{
\begin{tikzpicture}[scale=1]
	\node[draw,shape=circle,fill,scale=.5] (C) at (-30:1.9) {} ;
	\node[draw,shape=circle,fill,scale=.5] (B) at (90:1.9) {} ;
    \node[draw,shape=circle,fill,scale=.5]  (A) at (210:1.9) {} ;
    \node[draw,shape=circle,fill,scale=.5]  (c) at ($(A)!0.5!(B)$) {} ;
    \node[draw,shape=circle,fill,scale=.5]  (b) at ($(A)!0.5!(C)$) {} ;
    \node[draw,shape=circle,fill,scale=.5]  (a) at ($(C)!0.5!(B)$) {} ;
    \draw[thick,gray] plot [smooth cycle,tension=.8] coordinates { ($(B)+(90:0.2)$) ($(a)+(-15:0.2)$) ($(c)+(195:0.2)$) } ;
    \draw[thick,gray] plot [smooth cycle,tension=.8] coordinates { ($(c)+(90:0.2)$) ($(b)+(-15:0.2)$) ($(A)+(195:0.2)$) } ;
    \draw[thick,gray] plot [smooth cycle,tension=.8] coordinates { ($(a)+(90:0.2)$) ($(C)+(-15:0.2)$) ($(b)+(195:0.2)$) } ;

    \node at (-30:2.5) {$C$};
    \node at (90:2.5) {$B$};
    \node at (210:2.5) {$A$};
    \node at (30:1.5) {$a$};
    \node at (150:1.5) {$c$};
    \node at (270:1.5) {$b$};
\end{tikzpicture}
}    
\subfigure[Scenario $H_2$]{
\begin{tikzpicture}[scale=1]
	\node[draw,shape=circle,fill,scale=.5] (C) at (-30:1.9) {} ;
	\node[draw,shape=circle,fill,scale=.5] (B) at (90:1.9) {} ;
    \node[draw,shape=circle,fill,scale=.5]  (A) at (210:1.9) {} ;
    \node[draw,shape=circle,fill,scale=.5]  (c) at ($(A)!0.5!(B)$) {} ;
    \node[draw,shape=circle,fill,scale=.5]  (b) at ($(A)!0.5!(C)$) {} ;
    \node[draw,shape=circle,fill,scale=.5]  (a) at ($(C)!0.5!(B)$) {} ;
    \draw[thick,gray] plot [smooth cycle,tension=.8] coordinates { ($(B)+(90:0.2)$) ($(a)+(-15:0.2)$) ($(c)+(195:0.2)$) } ;
    \draw[thick,gray] plot [smooth cycle,tension=.8] coordinates { ($(c)+(90:0.2)$) ($(b)+(-15:0.2)$) ($(A)+(195:0.2)$) } ;
    \draw[thick,gray] plot [smooth cycle,tension=.8] coordinates { ($(a)+(90:0.2)$) ($(C)+(-15:0.2)$) ($(b)+(195:0.2)$) } ;
    \draw[thick,darkgreen] plot [smooth cycle,tension=.8] coordinates { ($(b)+(270:0.2)$) ($(a)+(15:0.2)$) ($(c)+(165:0.2)$) } ;

    \node at (-30:2.5) {$C$};
    \node at (90:2.5) {$B$};
    \node at (210:2.5) {$A$};
    \node at (30:1.5) {$a$};
    \node at (150:1.5) {$c$};
    \node at (270:1.5) {$b$};
\end{tikzpicture}
}      
\caption{Scenarios $H_1$ and $H_2$ used in Prop.~\ref{prop:6}. Notice that $H_2$ has an extra hyperedge ($\{a,b,c\}$, depicted in green) but is otherwise identical to $H_1$.}
\label{fig:mareada}
\end{figure}

\begin{prop}\label{prop:6}
    Consider the contextuality scenarios $H_1$ and $H_2$, depicted in Fig.~\ref{fig:mareada}, with vertex sets
\begin{align}
    V(H_1)=V(H_2)=\{A,B,C,a,b,c\}\,, \label{eq:verse3}
\end{align}
and hyperedge sets
\begin{align}
    E(H_1) &= \{\{A,b,c\}, \{a,B,c\},\{a,b,C\}\}\,,\\
    E(H_2) &= \{\{A,b,c\}, \{a,B,c\},\{a,b,C\},\{a,b,c\}\}\,.
\end{align}
These scenarios satisfy the following: 
\begin{align} 
    \mathrm{OrthoGraph}(H_1) &= \mathrm{OrthoGraph}(H_2)\,,\\
    \overbar{H}_1 &\neq \overbar{H}_2\,, \label{eq:neqcom3}\\
    \mathcal{Q}_1(H_1) &\neq \mathcal{Q}_1(H_2)\,. \label{eq:neqaq3}
\end{align}
\end{prop}
\begin{proof}
First, notice that the extra hyperedge $\{a,b,c\}$ imposes no additional orthogonality constraints than what's already covered by the hyperedges in $H_1$. Hence, $\mathrm{OrthoGraph}(H_1) = \mathrm{OrthoGraph}(H_2)$. 

Now, notice that the following is a valid behaviour in $H_1$:
\begin{align}
    \mathbf{p} = \left[\frac{1}{2},\frac{1}{2},\frac{1}{2},\frac{1}{4},\frac{1}{4},\frac{1}{4} \right]\,,
\end{align}
where the vertices are presented in the order of Eq.~\eqref{eq:verse3}. Moreover, this is a quantum behaviour in $H_1$ realised by the following state and projectors: 
\begin{align}
    &\ket{\Psi} = (\ket{1}+\ket{2}+\ket{3}+\ket{4})/2\,,\\
    &P_a = \ket{1}\bra{1}\,,\quad P_b = \ket{2}\bra{2}\,,\quad P_c = \ket{3}\bra{3}\,,\\
    &P_C = \ket{3}\bra{3}+\ket{4}\bra{4} \,,\quad P_A = \ket{1}\bra{1}+\ket{4}\bra{4} \,,\quad P_B = \ket{2}\bra{2}+\ket{4}\bra{4}\,.
\end{align}
Hence, $\mathbf{p} \in \mathcal{Q}_1(H_1)$. However, $p(a)+p(b)+p(c) = \frac{3}{4} \neq 1$. Hence $\mathbf{p} \not\in \mathcal{G}(H_2)$ and so $\mathbf{p} \not\in \mathcal{Q}_1(H_2)$. This proves Eq.~\eqref{eq:neqaq3}.

Now we need to show that $\overbar{H}_1 \neq \overbar{H}_2$. This follows from the fact that the previously-defined $\mathbf{p}$ implies $\mathcal{G}(H_1) \neq \mathcal{G}(H_2)$, which means that $\overbar{H}_1 = \overbar{H}_2$ cannot be true. 
\end{proof}

\bigskip

The previous example shows that equal orthogonality graphs is \textit{not sufficient} for the almost-quantum sets of two scenarios to be the same. Next, we will see that equal orthogonality graphs is also \textit{not necessary} for the almost-quantum sets of two scenarios to coincide. This is true even when the scenarios have different completions. 

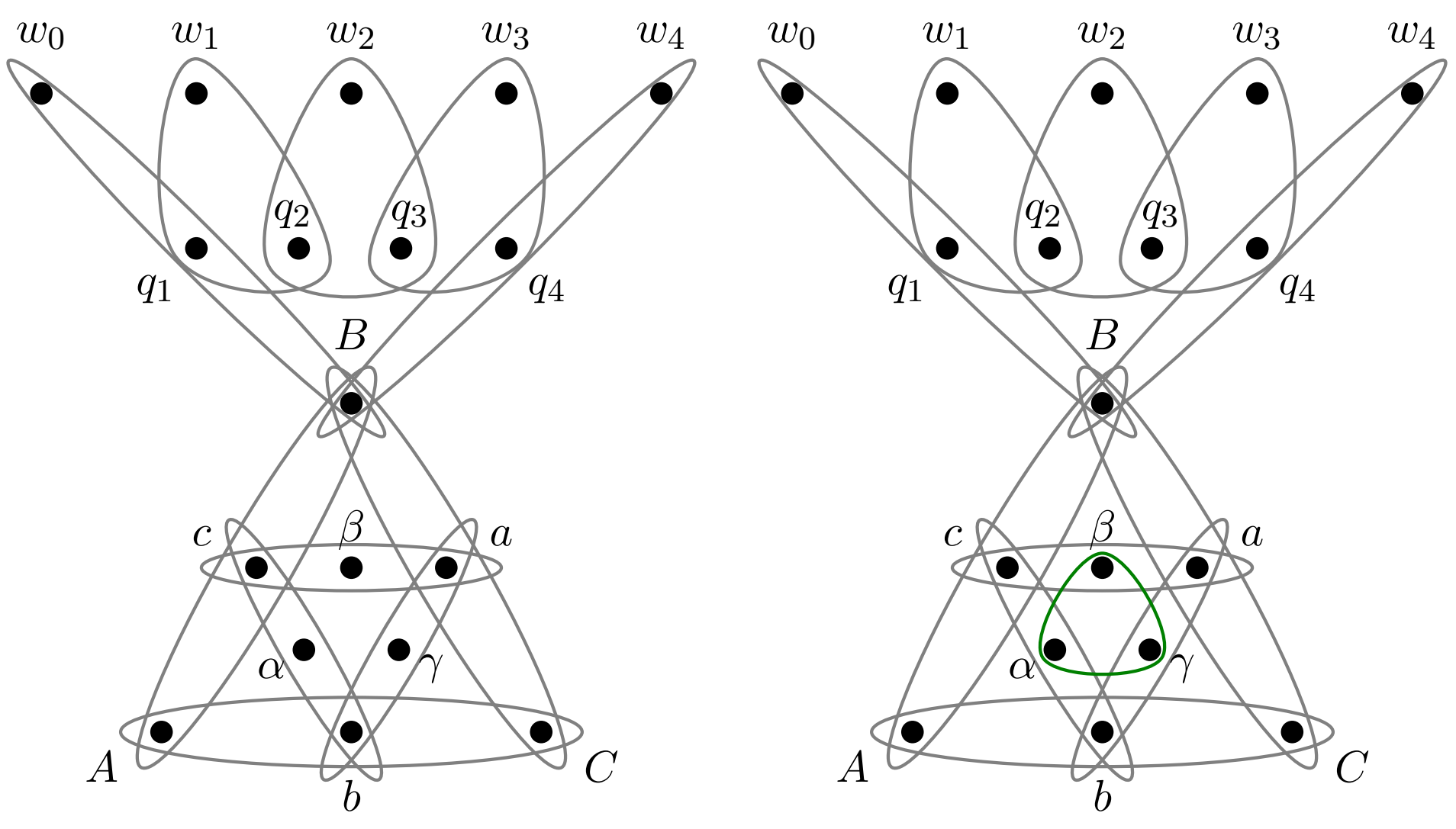
\begin{figure}
\centering
\subfigure[Scenario $H_1$]{
\begin{tikzpicture}[scale=1]

	\node[draw,shape=circle,fill,scale=.5] (C) at (-30:1.9) {} ;
	\node[draw,shape=circle,fill,scale=.5] (B) at (90:1.9) {} ;
    \node[draw,shape=circle,fill,scale=.5]  (A) at (210:1.9) {} ;
    \node[draw,shape=circle,fill,scale=.5]  (c) at ($(A)!0.5!(B)$) {} ;
    \node[draw,shape=circle,fill,scale=.5]  (b) at ($(A)!0.5!(C)$) {} ;
    \node[draw,shape=circle,fill,scale=.5]  (a) at ($(C)!0.5!(B)$) {} ;
    \node[draw,shape=circle,fill,scale=.5]  (alpha) at ($(b)!0.5!(c)$) {} ;
    \node[draw,shape=circle,fill,scale=.5]  (beta) at ($(a)!0.5!(c)$) {} ;
    \node[draw,shape=circle,fill,scale=.5]  (gamma) at ($(b)!0.5!(a)$) {} ;

    \node[draw,shape=circle,fill,scale=.5] (q1) at ($(B)+(135:1.9)$) {};
    \node[draw,shape=circle,fill,scale=.5] (q4) at ($(B)+(45:1.9)$) {};
    \node[draw,shape=circle,fill,scale=.5] (w0) at ($(q1)+(135:1.9)$) {};
    \node[draw,shape=circle,fill,scale=.5] (w4) at ($(q4)+(45:1.9)$) {};
    \node[draw,shape=circle,fill,scale=.5] (q2) at ($(q1)!0.33!(q4)$) {};
    \node[draw,shape=circle,fill,scale=.5] (q3) at ($(q1)!0.66!(q4)$) {};

    \node[draw,shape=circle,fill,scale=.5] (w1) at ($(w0)!0.25!(w4)$) {};
    \node[draw,shape=circle,fill,scale=.5] (w2) at ($(w0)!0.5!(w4)$) {};
    \node[draw,shape=circle,fill,scale=.5] (w3) at ($(w0)!0.75!(w4)$) {};

    \draw[thick,gray, rotate around={60:(c)}] (c) ellipse (2cm and .3cm) ;
    \draw[thick,gray, rotate around={120:(a)}] (a) ellipse (2cm and .3cm) ;
    \draw[thick,gray, rotate around={0:(b)}] (b) ellipse (2cm and .3cm) ;

    \draw[thick,gray, rotate around={60:(gamma)}] (gamma) ellipse (1.3cm and .2cm) ;
    \draw[thick,gray, rotate around={120:(alpha)}] (alpha) ellipse (1.3cm and .2cm) ;
    \draw[thick,gray, rotate around={0:(beta)}] (beta) ellipse (1.3cm and .2cm) ;

    \draw[thick,gray, rotate around={45:(q4)}] (q4) ellipse (2.3cm and .2cm) ;
    \draw[thick,gray, rotate around={135:(q1)}] (q1) ellipse (2.3cm and .2cm) ;

    \draw[thick,gray] plot [smooth cycle,tension=.8] coordinates { ($(w1)+(90:0.3)$) ($(q1)+(195:0.2)$) ($(q2)+(-25:0.3)$) } ;

\draw[thick,gray] plot [smooth cycle,tension=.8] coordinates { ($(w2)+(90:0.3)$) ($(q2)+(205:0.3)$) ($(q3)+(-25:0.3)$) } ;

\draw[thick,gray] plot [smooth cycle,tension=.8] coordinates { ($(w3)+(90:0.3)$) ($(q3)+(205:0.3)$) ($(q4)+(-15:0.2)$) } ;


    \node at (-30:2.5) {$C$};
    \node at (90:2.5) {$B$};
    \node at (210:2.5) {$A$};

    \node at (30:1.5) {$a$};
    \node at (150:1.5) {$c$};
    \node at (270:1.5) {$b$};

    \node at (-30:0.8) {$\gamma$};
    \node at (90:0.8) {$\beta$};
    \node at (210:0.8) {$\alpha$};

    \node at ($(q1)+(225:0.5)$) {$q_1$};
    \node at ($(q4)+(-45:0.5)$) {$q_4$};
    \node at ($(q2)+(100:0.3)$) {$q_2$};
    \node at ($(q3)+(75:0.3)$) {$q_3$};

    \node at ($(w0)+(90:0.5)$) {$w_0$};
    \node at ($(w1)+(90:0.5)$) {$w_1$};
    \node at ($(w2)+(90:0.5)$) {$w_2$};
    \node at ($(w3)+(90:0.5)$) {$w_3$};
    \node at ($(w4)+(90:0.5)$) {$w_4$};

\end{tikzpicture}
}    
\subfigure[Scenario $H_2$]{
\begin{tikzpicture}[scale=1]

	\node[draw,shape=circle,fill,scale=.5] (C) at (-30:1.9) {} ;
	\node[draw,shape=circle,fill,scale=.5] (B) at (90:1.9) {} ;
    \node[draw,shape=circle,fill,scale=.5]  (A) at (210:1.9) {} ;
    \node[draw,shape=circle,fill,scale=.5]  (c) at ($(A)!0.5!(B)$) {} ;
    \node[draw,shape=circle,fill,scale=.5]  (b) at ($(A)!0.5!(C)$) {} ;
    \node[draw,shape=circle,fill,scale=.5]  (a) at ($(C)!0.5!(B)$) {} ;
    \node[draw,shape=circle,fill,scale=.5]  (alpha) at ($(b)!0.5!(c)$) {} ;
    \node[draw,shape=circle,fill,scale=.5]  (beta) at ($(a)!0.5!(c)$) {} ;
    \node[draw,shape=circle,fill,scale=.5]  (gamma) at ($(b)!0.5!(a)$) {} ;

    \node[draw,shape=circle,fill,scale=.5] (q1) at ($(B)+(135:1.9)$) {};
    \node[draw,shape=circle,fill,scale=.5] (q4) at ($(B)+(45:1.9)$) {};
    \node[draw,shape=circle,fill,scale=.5] (w0) at ($(q1)+(135:1.9)$) {};
    \node[draw,shape=circle,fill,scale=.5] (w4) at ($(q4)+(45:1.9)$) {};
    \node[draw,shape=circle,fill,scale=.5] (q2) at ($(q1)!0.33!(q4)$) {};
    \node[draw,shape=circle,fill,scale=.5] (q3) at ($(q1)!0.66!(q4)$) {};

    \node[draw,shape=circle,fill,scale=.5] (w1) at ($(w0)!0.25!(w4)$) {};
    \node[draw,shape=circle,fill,scale=.5] (w2) at ($(w0)!0.5!(w4)$) {};
    \node[draw,shape=circle,fill,scale=.5] (w3) at ($(w0)!0.75!(w4)$) {};

    \draw[thick,gray, rotate around={60:(c)}] (c) ellipse (2cm and .3cm) ;
    \draw[thick,gray, rotate around={120:(a)}] (a) ellipse (2cm and .3cm) ;
    \draw[thick,gray, rotate around={0:(b)}] (b) ellipse (2cm and .3cm) ;

    \draw[thick,gray, rotate around={60:(gamma)}] (gamma) ellipse (1.3cm and .2cm) ;
    \draw[thick,gray, rotate around={120:(alpha)}] (alpha) ellipse (1.3cm and .2cm) ;
    \draw[thick,gray, rotate around={0:(beta)}] (beta) ellipse (1.3cm and .2cm) ;

    \draw[thick,gray, rotate around={45:(q4)}] (q4) ellipse (2.3cm and .2cm) ;
    \draw[thick,gray, rotate around={135:(q1)}] (q1) ellipse (2.3cm and .2cm) ;

    \draw[thick,gray] plot [smooth cycle,tension=.8] coordinates { ($(w1)+(90:0.3)$) ($(q1)+(195:0.2)$) ($(q2)+(-25:0.3)$) } ;

\draw[thick,gray] plot [smooth cycle,tension=.8] coordinates { ($(w2)+(90:0.3)$) ($(q2)+(205:0.3)$) ($(q3)+(-25:0.3)$) } ;

\draw[thick,gray] plot [smooth cycle,tension=.8] coordinates { ($(w3)+(90:0.3)$) ($(q3)+(205:0.3)$) ($(q4)+(-15:0.2)$) } ;

\draw[thick,darkgreen] plot [smooth cycle,tension=.8] coordinates { (-30:0.6) (90:0.6) (210:0.6) } ;

    \node at (-30:2.5) {$C$};
    \node at (90:2.5) {$B$};
    \node at (210:2.5) {$A$};

    \node at (30:1.5) {$a$};
    \node at (150:1.5) {$c$};
    \node at (270:1.5) {$b$};

    \node at (-30:0.8) {$\gamma$};
    \node at (90:0.8) {$\beta$};
    \node at (210:0.8) {$\alpha$};

    \node at ($(q1)+(225:0.5)$) {$q_1$};
    \node at ($(q4)+(-45:0.5)$) {$q_4$};
    \node at ($(q2)+(100:0.3)$) {$q_2$};
    \node at ($(q3)+(75:0.3)$) {$q_3$};

    \node at ($(w0)+(90:0.5)$) {$w_0$};
    \node at ($(w1)+(90:0.5)$) {$w_1$};
    \node at ($(w2)+(90:0.5)$) {$w_2$};
    \node at ($(w3)+(90:0.5)$) {$w_3$};
    \node at ($(w4)+(90:0.5)$) {$w_4$};

\end{tikzpicture}
}      
\caption{Scenarios $H_1$ and $H_2$ used in Prop.~\ref{pr:ex18} to provide a counterexample to Conjecture \#1 of Ref.~\cite{sainz2018multipartite} where the scenarios have different completions. Notice that $H_2$ has an extra hyperedge ($\{\alpha,\beta,\gamma\}$, depicted in green) but is otherwise identical to $H_1$.}
\label{fig:aqc3}
\end{figure}

\begin{prop}\label{pr:ex18}
Consider the contextuality scenarios $H_1$ and $H_2$, depicted in Fig.~\ref{fig:aqc3}, with vertex sets
\begin{align}
    V(H_1)=V(H_2)=\{A,B,C,a,b,c,\alpha,\beta,\gamma,q_1,q_2,q_3,q_4,w_0,w_1,w_2,w_3,w_4\}\,,  \label{eq:ver18}
\end{align}
and hyperedge sets
\begin{align}
    E(H_1) =  \{&\{A,B,c\}, \{a,B,C\},\{A,b,C\},\{a,\beta,c\},\{\alpha,b,c\},\{a,b,\gamma\}, \\ \nonumber &\{B,q_1,w_0\},\{q_1,q_2,w_1\}, \{q_2,q_3,w_2\},\{q_3,q_4,w_3\},\{B,q_4,w_4\} \}\,,\\
    E(H_2) =  \{&\{A,B,c\}, \{a,B,C\},\{A,b,C\},\{a,\beta,c\},\{\alpha,b,c\},\{a,b,\gamma\},\\ \nonumber 
    &\{B,q_1,w_0\},\{q_1,q_2,w_1\}, \{q_2,q_3,w_2\},\{q_3,q_4,w_3\},\{B,q_4,w_4\},\{\alpha,\beta,\gamma\} \} \,.
\end{align}
These scenarios satisfy the following: 
\begin{align} \label{eq:neqor2}
    \mathrm{OrthoGraph}(H_1) &\neq \mathrm{OrthoGraph}(H_2)\,,\\
    \overbar{H}_1 &\neq \overbar{H}_2\,, \label{eq:neqcom2}\\
    \mathcal{Q}_1(H_1) &= \mathcal{Q}_1(H_2)\,,\label{eq:eqaq2} \\
    \mathcal{C}(H_i) \neq &\mathcal{Q}_1(H_i) \neq \mathcal{G}(H_i)\,, \quad i=1,2. \label{eq:hiernocol}
\end{align}
\end{prop}
The proof of Prop.~\ref{pr:ex18} is presented in Appendix \ref{ap:proofex}. Simpler examples can also be found, if one does not demand Eq.~\eqref{eq:hiernocol} (see Props.~\ref{pr:ex9} and \ref{pr:aqc} in Appendix \ref{ap:proofex}) and Eq.~\eqref{eq:neqcom2} (see Prop.~\ref{pr:aqc} in Appendix \ref{ap:proofex}). Notice then that Props.~\ref{pr:ex18}, \ref{pr:ex9}, and \ref{pr:aqc} are counterexamples to Conjecture [\# 1 of Ref.~\cite{sainz2018multipartite}], and hence we have disproven this conjecture. 

\bigskip

We have seen then that having equal orthogonality graphs is neither a necessary nor a sufficient condition for two contextuality scenarios to admit of the same set of almost-quantum behaviours. We conclude this section by showing how to complement the condition $\mathrm{OrthoGraph}(H_1) = \mathrm{OrthoGraph}(H_2)$ for the almost-quantum sets to be the same. 

\bigskip

\begin{prop}\label{pr:eOeGeAQ}
Let $H_1$ and $H_2$ be two contextuality scenarios with the following properties: 
\begin{align}
     \mathrm{OrthoGraph}(H_1) &= \mathrm{OrthoGraph}(H_2)   \,,\\
     \mathcal{G}(H_1) &= \mathcal{G}(H_2)\,.
\end{align}
Then, 
\begin{align}
    \mathcal{Q}_1(H_1) = \mathcal{Q}_1(H_2)\,.
\end{align}
\end{prop}
\begin{proof}
First notice that $\mathrm{OrthoGraph}(H_1) = \mathrm{OrthoGraph}(H_2)$ implies that $\mathrm{NO}(H_1) = \mathrm{NO}(H_2)$. Now, Ref.~\cite[Prop.~6.3.2]{acin2015combinatorial} shows that 
$\mathbf{p} \in \mathcal{Q}_1(H) \,\, \Leftrightarrow \,\, \mathbf{p} \in \mathcal{G}(H) \,\, \text{and} \,\, \vartheta(\mathrm{NO}(H),\mathbf{p}) = 1$, for any $H$. Hence,  
\begin{align}
    \mathbf{p}\in\mathcal{Q}_1(H_1) &\Leftrightarrow \,\, \mathbf{p} \in \mathcal{G}(H_1)\,,\, \vartheta(\mathrm{NO}(H_1),\mathbf{p})=1 \,,\\ \label{eq:themid}
    &\Leftrightarrow \,\, \mathbf{p} \in \mathcal{G}(H_2)\,,\, \vartheta(\mathrm{NO}(H_2),\mathbf{p})=1 \,,\\
    &\Leftrightarrow \,\, \mathbf{p}\in\mathcal{Q}_1(H_2)\,, 
\end{align}
where Eq.~\eqref{eq:themid} follows from $\mathcal{G}(H_1) = \mathcal{G}(H_2)$ and $\mathrm{NO}(H_1) = \mathrm{NO}(H_2)$. 

Therefore, $\mathcal{Q}_1(H_1) = \mathcal{Q}_1(H_2)$. 
\end{proof}

A corollary of Prop.~\ref{pr:eOeGeAQ} is the following. 

\begin{cor}
Let $H_1$ and $H_2$ be two contextuality scenarios with the following properties: 
$\mathrm{OrthoGraph}(H_1) = \mathrm{OrthoGraph}(H_2)$ and $\overbar{H}_1 = \overbar{H}_2$. Then, $\mathcal{Q}_1(H_1) = \mathcal{Q}_1(H_2)$.
\end{cor}
\begin{proof}
First notice that $\overbar{H}_1 = \overbar{H}_2$ implies that $\mathcal{G}(H_1) = \mathcal{G}(\overbar{H}_1) = \mathcal{G}(\overbar{H}_2) =    \mathcal{G}(H_1)$. 
Hence $\mathcal{G}(H_1) = \mathcal{G}(H_2)$, and so the claim follows from Prop.~\ref{pr:eOeGeAQ}.
\end{proof}
We notice that this Corollary also follow from Ref.~\cite[Prop.~C.1.6(b)]{acin2015combinatorial}. 

\bigskip

Let us conclude this section with a corollary from Props.~\ref{pr:ex18}, \ref{pr:ex9}, and \ref{pr:aqc} that pertains to properties of the Lov\'asz number \cite{lovasz}. 

\begin{cor}
There exist hypergraphs $H_1$ and $H_2$ such that $\mathrm{NO}(H_1) \neq \mathrm{NO}(H_2)$, but nonetheless for all $\mathbf{p} \in \mathcal{G}(H_1) \cap \mathcal{G}(H_2)$
\begin{align}
    \vartheta(\mathrm{NO}(H_1),\mathbf{p}) = 1 \quad \text{iff} \quad  \vartheta(\mathrm{NO}(H_2),\mathbf{p})=1\,.
\end{align}
\end{cor}
\begin{proof}
    This follows equally   from Prop.~\ref{pr:ex18}, Prop.~\ref{pr:ex9}, or Prop.~\ref{pr:aqc}, after recalling that ${\mathbf{p} \in \mathcal{Q}_1(H)}$ iff $\vartheta(\mathrm{NO}(H),\mathbf{p}) =1$ \cite[Prop.~6.3.2]{acin2015combinatorial}.
\end{proof}

\section{Quantum vs.~almost-quantum behaviours}\label{se:aqdiff}

In this section we present necessary and sufficient conditions for two scenarios to display the same sets of quantum or almost-quantum behaviours. These conditions differ depending on whether one asks about quantum or almost-quantum, and hence pose grounds on which to explore a principled difference between these two. 

\bigskip

We begin by defining a property of sets of behaviours, which we call \textit{mutual normalisation} and denote by $\MN$. 

\begin{defn}\label{def:MN} ($\, \MN \, $) \\
Consider two contextuality scenarios $H_1$ and $H_2$. Let $\mathcal{S}_1 \subseteq \mathcal{G}(H_1)$ and $\mathcal{S}_2 \subseteq \mathcal{G}(H_2)$ be sets of behaviours in $H_1$ and $H_2$. Then $\mathcal{S}_1$ and $\mathcal{S}_2$ satisfy $\MN$ iff
\begin{align}\label{eq:MN1}
    &\sum_{v \in e} p(v)=1 \quad \forall \, e \in E(H_2) \,,\quad \forall \, \mathbf{p} \in \mathcal{S}_1\,,\\ \label{eq:MN2}
    &\sum_{v \in e} p(v)=1 \quad \forall \, e \in E(H_1) \,,\quad \forall \, \mathbf{p} \in \mathcal{S}_2\,.
\end{align}
\end{defn}

First, we notice how the essence of quantum behaviours may be captured by $\MN$ without the need to consider structural relations between the hypergraphs.  

\begin{prop}\label{prop:necsufquan}
    Consider two contextuality scenarios $H_1$ and $H_2$ with the same vertex set. Their quantum sets of behaviours $\mathcal{Q}(H_1)$ and $\mathcal{Q}(H_2)$ satisfy $\MN$ if and only if $\mathcal{Q}(H_1) = \mathcal{Q}(H_2)$.
\end{prop}
\begin{proof}
Suppose that $\mathcal{Q}(H_1) = \mathcal{Q}(H_2)$. On the one hand, this implies $\mathcal{Q}(H_1) \subseteq \mathcal{G}(H_2)$, which implies Eq.~\eqref{eq:MN1} for $\mathcal{S}_1 = \mathcal{Q}(H_1)$. And on the other hand, it also implies  $\mathcal{Q}(H_2) \subseteq \mathcal{G}(H_1)$, which implies Eq.~\eqref{eq:MN2} for $\mathcal{S}_2 = \mathcal{Q}(H_2)$. Hence, $\mathcal{Q}(H_1)$ and $\mathcal{Q}(H_2)$ satisfy $\MN$.

For the converse, let us assume that Eqs.~\eqref{eq:MN1} and \eqref{eq:MN2} hold when $\mathcal{S}_k = \mathcal{Q}(H_k)$ with $k=1,2$. Consider an arbitrary quantum behaviour in $H_1$: $\mathbf{p} \in \mathcal{Q}(H_1)$. From Ref.~\cite[Definition 5.1.1]{acin2015combinatorial} there exists a Hilbert space $\mathcal{H}$, a density operator $\rho$ in this space, and projections $\{P_v\}_{v \in V(H_1)}$ in $\mathcal{H}$ such that 
\begin{align} \label{eq:projnormq}
    \sum_{v \in e} P_v &= \id \quad \forall \, e \in E(H_1)\,,\\ \label{eq:probcompq}
    p(v) &= \tr{P_v \, \rho}{}\quad \forall \, v \in V(H_1)\,.
\end{align}
We will now show that this set of projectors also satisfies $\sum_{v \in e} P_v = \id \quad \forall \, e \in E(H_2)$ and hence, together with $\rho$, they define a valid quantum behaviour in $H_2$. Consider then an arbitrary unit vector $\ket{\psi} \in \mathcal{H}$. Together with the projectors $\{P_v\}_{v \in V(H_1)}$ (since they satisfy Eq.~\eqref{eq:projnormq}) they define a quantum behaviour in $H_1$ with probability assignment $p_\psi(v)= \bra{\psi} P_v \ket{\psi}$\,. Eq.~\eqref{eq:MN1} then implies that for any $e \in E(H_2)$
\begin{align}
    \sum_{v \in e} \bra{\psi} P_v \ket{\psi} = 1\,. 
\end{align}
But recall that this is valid for any arbitrary state $\ket{\psi}$. Hence, $\sum_{v \in e} P_v  = \id$ for every $e \in E(H_2)$. From this follows that $\mathbf{p} \in \mathcal{Q}(H_2)$, which holds for any quantum behaviour in $H_1$. A similar argument starting from an arbitrary quantum behaviour $\mathbf{p}'$ in $H_2$ yields $\mathbf{p}' \in \mathcal{Q}(H_1)$. Hence, $\mathcal{Q}(H_1)=\mathcal{Q}(H_2)$, which concludes the proof.     
\end{proof}

As we discuss in Appendix \ref{ap:algo}, checking whether two scenarios admit the same set of quantum behaviours is an undecidable problem. 

\bigskip

For almost-quantum behaviours, however, $\MN$ alone cannot single out their essence, as we see in the following propositions. 

\begin{prop}\label{prop:necsufeo}
Let $H_1$ and $H_2$ be two contextuality scenarios with the same vertex set, and which satisfy  $\mathrm{OrthoGraph}(H_1) = \mathrm{OrthoGraph}(H_2)   $. 

Then $\mathcal{Q}_1(H_1) = \mathcal{Q}_1(H_2)$ if and only if $\mathcal{Q}_1(H_1)$ and $\mathcal{Q}_1(H_2)$ satisfy $\MN$.
\end{prop}

\begin{proof}
Let us start with the `only if' direction. If $\mathcal{Q}_1(H_1) = \mathcal{Q}_1(H_2)$, then $\mathbf{p}\in \mathcal{Q}_1(H_1) \,\, \Leftrightarrow \,\, \mathbf{p}\in \mathcal{Q}_1(H_2)$ and so Eq.~\eqref{eq:MN1} holds. In addition,     if $\mathcal{Q}_1(H_1) = \mathcal{Q}_1(H_2)$, then $\mathbf{p}\in \mathcal{Q}_1(H_2) \,\, \Leftrightarrow \,\, \mathbf{p}\in \mathcal{Q}_1(H_1)$ and so Eq.~\eqref{eq:MN2} holds.

Now let us show the `if' part. Take $\mathbf{p}\in \mathcal{Q}_1(H_1)$. Eq.~\eqref{eq:MN1} implies that $\mathbf{p}\in \mathcal{G}(H_2)$. Since $\mathbf{p}\in \mathcal{Q}_1(H_1)$, then $\vartheta(\mathrm{NO}(H_1),\mathbf{p}) = 1$. Since  $\mathrm{OrthoGraph}(H_1) = \mathrm{OrthoGraph}(H_2)   $ then $\vartheta(\mathrm{NO}(H_2),\mathbf{p}) = 1$. Since $\mathbf{p}\in \mathcal{G}(H_2)$, then Ref.~\cite[Prop.~6.3.2]{acin2015combinatorial} implies that $\mathbf{p}\in \mathcal{Q}_1(H_2)$, and hence $\mathcal{Q}_1(H_1) \subseteq \mathcal{Q}_1(H_2)$. Now, the same argument but with the roles of $H_1$ and $H_2$ interchanged shows that  $\mathcal{Q}_1(H_2) \subseteq \mathcal{Q}_1(H_1)$. Hence, $\mathcal{Q}_1(H_1) = \mathcal{Q}_1(H_2)$. 
\end{proof}

In Prop.~\ref{pr:sdpMNeo} of Appendix \ref{ap:algo} we show how $\mathcal{Q}_1(H_1) = \mathcal{Q}_1(H_2)$ can be certified with two semidefinite programs when $\mathrm{OrthoGraph}(H_1) = \mathrm{OrthoGraph}(H_2)$. 

\begin{prop}
There exist contextuality scenarios $H_1$ and $H_2$ with the same vertex set and different orthogonality graphs, whose sets of almost-quantum behaviours satisfy $\MN$, but for which $\mathcal{Q}_1(H_1)\neq\mathcal{Q}_1(H_2)$.
\end{prop}
\begin{proof}
Let us present an example of two contextuality scenarios $H_1$ and $H_2$ such that:  $\mathcal{Q}_1(H_1)$ and $\mathcal{Q}_1(H_2)$ satisfy $\MN$, $\mathrm{OrthoGraph}(H_1) \neq \mathrm{OrthoGraph}(H_2)$, and $\mathcal{Q}_1(H_1) \neq \mathcal{Q}_1(H_2)$. These scenarios are the ones defined in Ref.~\cite[Appendix E]{sainz2018multipartite}, and we will briefly include them here. 

Scenarios $H_1$ and $H_2$ have the same vertex sets of 16 elements. Scenario $H_1$ has 8 hyperedges $E_1 = \{e_1\,,\, \ldots \,,\, e_8\}$, while scenario $H_2$ has 12 hyperedges $E_2 = E_1 \cup \{e_9\,,\, \ldots \,,\, e_{12}\}$, where the $\{e_k\}_{k=1:12}$ are defined in Ref.~\cite[Appendix E]{sainz2018multipartite}. These scenarios satisfy $\mathcal{G}(H_1)=\mathcal{G}(H_2)$ and hence Eqs.~\eqref{eq:MN1} and \eqref{eq:MN2} hold. However, $\mathcal{Q}_1(H_1) \neq \mathcal{Q}_1(H_2)$ as was shown in Ref.~\cite[Appendix E]{sainz2018multipartite} by finding an inequality that both sets give different maximum violations to. 
\end{proof}

These propositions illustrate how structural aspects of the hypergraphs themselves are required to capture almost-quantumness in a way that are not needed for quantum behaviours. This highlights another way in which quantum and almost-quantum phenomena differ. 

From Section \ref{se:motex}, however, we learned that contextuality scenarios may admit of the same set of almost-quantum behaviours even if their orthogonality graphs differ. Hence, the requirement in Prop.~\ref{prop:necsufeo} that the two orthogonality graphs be the same could potentially be too demanding. In what remains of this section we discuss how to complement the condition $\MN$ to have two contextuality scenarios admit the same sets of almost-quantum behaviours. 

\bigskip

The property we define next pertains to orthonormal labellings of graphs, and we denote it by $\AOL$. It captures the idea that for a behaviour $\mathbf{p} \in \mathcal{Q}_1(H_1)$ (and similarly swapping the roles of $H_1$ and $H_2$) there exists an orthonormal labelling of $\mathrm{NO}(H_2)$ and a unit vector $\ket{\Psi}$ that together reproduce the probabilities of $\mathbf{p} \in \mathcal{Q}_1(H_1)$. Notice that this does not impose that $\mathbf{p} \in \mathcal{G}(H_2)$.

\begin{defn}\label{def:aol}($\, \AOL \,$) \\
Consider two contextuality scenarios $H_1$ and $H_2$ with the same vertex set. 
Their almost-quantum sets of behaviours $\mathcal{Q}_1(H_1)$ and $\mathcal{Q}_1(H_2)$ satisfy $\AOL$ if the following holds. 

For any $\mathbf{p} \in \mathcal{Q}_1(H_k)$ (with $k=1,2$), there exists a Hilbert space $\cH$, a normalised vector $\ket{\Psi}$ in $\cH$, and a set of projectors $\{P_v\}_{v\in V(H_k)}$ in $\cH$, such that
\begin{align} \label{eq:recprob}
    p(v) = \bra{\Psi} P_v \ket{\Psi}& \quad \forall \, v\in V(H_k)\,,\\
    u \perp_{H_{\neg k}} v \quad &\Rightarrow \quad P_u \perp P_v \,, \label{eq:30}
\end{align} 
where $\neg k \in \{1,2\}\setminus \{k\}$. Notice that the projectors $\{P_v\}_{v\in V}$ are not required to satisfy the orthogonality relations imposed by $H_k$.      
\end{defn}

As we discuss in Appendix \ref{ap:algo}, there is no easy algorithm to test the satisfaction of property $\AOL$ in general. However, property $\MN$ can be easily checked for almost-quantum behaviours through a finite set of semidefinite programs (see Prop.~\ref{pr:sdpMN}). 

\bigskip

The property of $\AOL$ together with $\MN$ then capture almost-quantum behaviours, as we see next. 

\begin{prop}\label{prop:necsuf}
Let $H_1$ and $H_2$ be two contextuality scenarios with the same vertex set. 
Then $\mathcal{Q}_1(H_1) = \mathcal{Q}_1(H_2)$ if and only if $\mathcal{Q}_1(H_1)$ and $\mathcal{Q}_1(H_2)$ satisfy $\MN$ and $\AOL$.
\end{prop}
\begin{proof}
Suppose $\mathcal{Q}_1(H_1) = \mathcal{Q}_1(H_2)$. Then,   $\mathcal{Q}_1(H_1) \subseteq \mathcal{G}(H_2)$  and $\mathcal{Q}_1(H_2) \subseteq \mathcal{G}(H_1)$, from which $\MN$ follows. In addition, take $\mathbf{p} \in \mathcal{Q}_1(H_k)$ ($k=1,2$). Since $\mathbf{p} \in \mathcal{Q}_1(H_{\neg k})$, there exists an almost-quantum model \cite[Prop.~6.3.1(d)]{acin2015combinatorial} that satisfies the orthogonality requirements of $H_{\neg k}$. These state and projectors provide the requirements in $\AOL$, and hence $\AOL$ follows.

Let us now prove the converse. Consider an arbitrary model $\mathbf{p} \in \mathcal{Q}_1(H_1)$.
Now take the state $\ket{\Psi}$ and projectors $\{P_v\}_{v\in V}$ from $\AOL$ which satisfy Eqs.~\eqref{eq:recprob} and \eqref{eq:30}. Since we assume $\MN$, Eq.~\eqref{eq:MN1} ensures that $\sum_{v\in e} \bra{\Psi} P_v \ket{\Psi} =1$ for all $e \in E(H_2)$, and hence $\mathbf{p} \in \mathcal{G}(H_2)$. The conditions that  $\ket{\Psi}$ and projectors $\{P_v\}_{v\in V}$ satisfy then provide an almost-quantum model for $\mathbf{p}$ in $H_2$ as per Ref.~\cite[Prop.~6.3.1(d)]{acin2015combinatorial}. Hence, $\mathbf{p} \in \mathcal{Q}_1(H_2)$. Since this holds for arbitrary $\mathbf{p}$ this implies $\mathcal{Q}_1(H_1) \subseteq \mathcal{Q}_1(H_2)$. A similar reasoning starting from $\mathbf{p} \in \mathcal{Q}_1(H_2)$ leads to $\mathcal{Q}_1(H_2) \subseteq \mathcal{Q}_1(H_1)$. From this follows that $\mathcal{Q}_1(H_1) = \mathcal{Q}_1(H_2)$, which concludes the proof. 

\end{proof}

From Props.~\ref{prop:necsufquan} and \ref{prop:necsuf} we then learn that for quantum behaviours normalisation of the probabilities along a set of vertices implies the orthogonality of the projectors in any underlying quantum model, whereas the orthogonality properties of the projectors underpinning an almost-quantum model do not follow from normalisation of the probabilities alone, and need to be imposed externally.

\section{Relations between the disjunctive and the completed FR-products}\label{se:counter}

In this section we explore some structural properties of contextuality scenarios. We focus on how the orthogonality and completion properties of hypergraphs behave under composition. 

Ref.~\cite{sainz2018multipartite} proposed a composition rule for hypergraphs based on the orthogonality relations among the global events, called disjunctive FR-product. In a nutshell, this product renders as hyperedges of the composed hypergraph any collection of vertices that are both ``mutually orthogonal'' and ``an edge or a virtual edge of the maximal FR-product of the individual scenarios''. In this sense, some virtual hyperedges of the composite system will not be considered hyperedges in ${}^{\disj}\otimes_i H_i$. When all virtual hyperedges of ${}^{\max}\otimes_i H_i$ are considered as real hyperedges of the composite system, one recovers the completed scenario $\overbar{\otimes}_i H_i$. 
Here we will explore how ${}^{\disj}\otimes$ and $\overbar{\otimes}$ relate, in particular for the case of contextuality scenarios that are self-complete. 

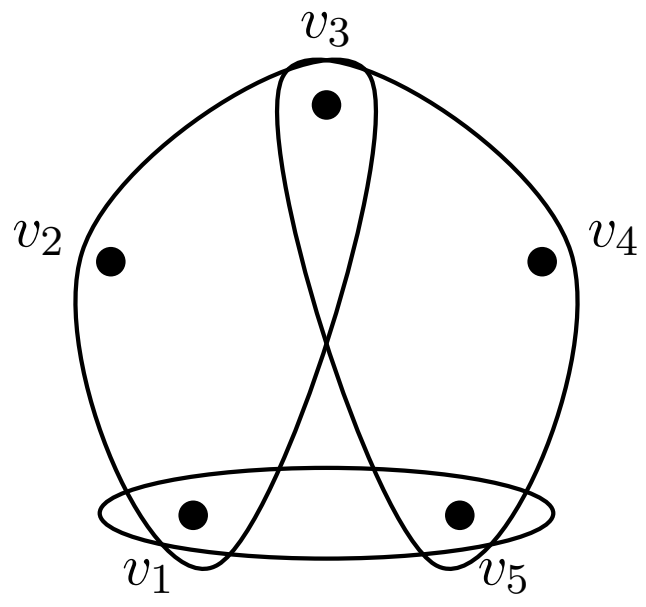
\begin{figure}
    \centering
    \begin{tikzpicture}[scale=1]

	\node[draw,shape=circle,fill,scale=.5] (a) at (18:1.5) {} ;
    \node[] at (18:2) {$v_4$} ;
    \node[draw,shape=circle,fill,scale=.5] (b) at (90:1.5) {} ;
    \node[] at (90:2) {$v_3$} ;
    \node[draw,shape=circle,fill,scale=.5] (c) at (162:1.5) {} ;
    \node[] at (162:2) {$v_2$} ;
    \node[draw,shape=circle,fill,scale=.5] (d) at (234:1.5) {} ;
    \node[] at (234:2) {$v_1$} ;
    \node[draw,shape=circle,fill,scale=.5] (e) at (306:1.5) {} ;
    \node[] at (306:2) {$v_5$} ;
    \draw[thick,black] plot [smooth cycle,tension=.8] coordinates { (245:1.7) (162:1.7) (80:1.7) } ;
    \draw[thick,black] plot [smooth cycle,tension=.8] coordinates { (295:1.7) (18:1.7) (100:1.7) } ;
    \draw[thick,black] (0,-1.2) ellipse (1.5cm and .3cm) ;

\node[draw,color=white] at (0,-2) {} ;
\end{tikzpicture}
    \caption{Contextuality scenario from Prop.~\ref{pr:no2} which gives a counter-example to Conjecture [\# 2 of Ref.~\cite{sainz2018multipartite}] and to Conjecture [Footnote 10 of Ref.~\cite{sainz2018multipartite}].}
    \label{fig:pentagon}
\end{figure}

\bigskip

Let us first consider the contextuality scenario of Fig.~\ref{fig:pentagon}. We will see next that this scenario is self-complete, and when taking two copies of itself different results are obtained when using ${}^{\disj}\otimes$ or $\overbar{\otimes}$. 

\begin{prop}\label{pr:no2}
The contextuality scenario $H=(V,E)$, depicted in Fig.~\ref{fig:pentagon}, with vertex set
\begin{align}
    V = \{v_1,v_2,v_3,v_4,v_5\}\,,
\end{align}
and hyperedge set
\begin{align}
    E = \{\{v_1,v_2,v_3\} , \{v_3,v_4,v_5\} , \{v_1,v_5\} \}
\end{align}
satisfies
\begin{align}
    \overbar{H} &= H\,,\\
    \overbar{H} \,  {}^{\disj}\otimes \overbar{H} & \neq \overbar{H} \,  \overbar{\otimes} \,  \overbar{H}\,.
\end{align}
\end{prop}
The proof of Prop.~\ref{pr:no2} is presented in Appendix \ref{ap:counter}. 

Notice that Prop.~\ref{pr:no2} hence disproves Conjecture [\# 2 of Ref.~\cite{sainz2018multipartite}]. Since $H$ is self-complete, this moreover shows how the disjunctive product fails to generically produce a self-complete composite hypergraph. The self-completeness of $H$ also implies that Prop.~\ref{pr:no2} disproves Conjecture [Footnote 10 of Ref.~\cite{sainz2018multipartite}], and hence the Foulis-Randall product of two scenarios also does not generically produce a self-complete composite hypergraph. 

\bigskip

Self-completeness alone is not sufficient for the disjunctive FR-product to become  the easy-to-compute tool with which to obtain the more-complex completed FR-product. Next we articulate a condition for individual contextuality scenarios so that their disjunctive FR-product actually yields the completed FR-product. 

\begin{lem}\label{lem:sc}
Consider a set of contextuality scenarios $\{H_i\}$, such that each satisfies the following property: for each pair of vertices $u,v \in V(H_i)$ such that $u \not\perp v$ (including $u=v$) there exists a behaviour $\mathbf{p} \in \mathcal{G}(H_i)$ such that $p(u)=1$ and $p(v)>0$. Then, the disjunctive product of these scenarios is self-complete: 
\begin{align}
    {}^\disj \otimes_i H_i = \overbar{{}^\disj \otimes_i H_i} =: \overbar{\otimes}_i H_i. 
\end{align}
\end{lem}
The proof of Lemma \ref{lem:sc} is presented in Appendix \ref{ap:counter}. Checking whether the contextuality scenarios $\{H_i\}$ satisfy the hypothesis of Lemma \ref{lem:sc} amounts to $|\{H_i\}|$ linear programs, as shown in Prop.~\ref{prop:LPlem} in Appendix \ref{ap:algo}. 

Finally, we notice that Lemma \ref{lem:sc} is a sufficient but not necessary condition for the disjunctive FR-product of the scenarios to be complete. This can be seen immediately from the scenario $H$ of Prop.~\ref{pr:no2} (see Fig.~\ref{fig:pentagon}). On the one hand, $H$ fails the hypothesis of Lemma \ref{lem:sc}, since Eq.~\eqref{eq:82} implies $p(v_3) \leq \frac{1}{2}$ for any behaviour $\mathbf{p} \in \mathcal{G}(H)$. However, the disjunctive product $H \, {}^\disj \otimes K$, with $K$ a contextuality scenario with a single dichotomic measurement, is self-complete.  

\section{Discussion}\label{se:disc}

In this paper we explored the connection between the almost-quantum behaviours admitted by a contextuality scenario, and its properties pertaining to completion and orthogonality graph. 
Indeed, we showed that the role orthogonality graphs play in the characterisation of almost-quantum behaviours is not as straightforward as previously thought.
We noticed that two scenarios that have the same orthogonality graph and the same set of general behaviours admit of the same set of almost-quantum behaviours. This conclusion also follows if the scenarios have the same orthogonality graphs and the same completions. However, having only the same orthogonality graph is not enough for the scenarios to have the same almost-quantum set. In turn, we showed that contextuality scenarios may admit the same set of almost-quantum behaviours, even if their orthogonality graphs are different. This disproves Conjecture \#1 of Ref.~\cite{sainz2018multipartite} and challenges our intuitions behind almost-quantum phenomena. Indeed, scenarios with the same completion and same general behaviour sets but different orthogonality graphs may admit the same sets of almost-quantum behaviours (see Prop.~\ref{pr:aqc}), which hints that the structural features required for almost-quantum models don't follow directly from the orthogonality graphs as previously thought. 

We then explored necessary and sufficient conditions for two contextuality scenarios to admit of the same sets of quantum behaviours and of the same sets of almost-quantum behaviours. Whereas for the case of quantum behaviours considerations only about probability normalisation were relevant, this was no longer the case for almost-quantum behaviours. For the latter, extra conditions pertaining to the orthogonality structure of the scenarios were necessary (although not the equality of their orthogonality graphs). We discussed algorithms to check for these normalisation and structural conditions, which reduce to two semidefinite programs when the scenarios have the same orthogonality graphs. 
The fact that the  necessary and sufficient conditions for two scenarios to admit the same set of quantum behaviours are different from those required for almost-quantum sets to be the same, motivates the following question: can this shed light on the search for physical principles that almost-quantum behaviours violate?

The results in this manuscript identify a further distinction between quantum and almost-quantum behaviours, but in terms of properties of the sharp measurements (the analogue of a projective measurement) that may realise them within a physical theory. Hence, the kind of principles that one may formulate will relate to conditions on sharp measurements in general theories. To explore this for almost-quantum, then, a prerequisite is to have some candidate operational or physical theory for almost-quantum phenomena, where the notions of measurements, effects, and sharpness are articulated. 
Principles that pertain to the measurement structure of a general theory have already been identified for hypothetical almost-quantum theories in previous works: measurements in an almost-quantum theory cannot satisfy the no-restriction hypothesis \cite{sainz2018almost} (that is, not every mathematically-admissible measurement is physically implementable in the theory), and sharp measurements in an almost-quantum theory must violate Specker's principle \cite{gonda2018almost} (that is, pair-wise compatible sharp measurements are not always all jointly measurable). 
Quantum theory, however, satisfies both the no-restriction hypothesis and Specker's principle. 
Our results here then highlight another property of sharp measurements in an almost-quantum theory, that differs from how sharp measurements in quantum theory behave: it cannot happen that the sharp effects associated to a collection of events whose probabilities are always normalised (despite belonging to different measurements) are always mutually orthogonal (that is, orthogonality relations of sharp effects cannot be read out from normalisation of the probabilities, unlike in quantum theory).

In this manuscript we have also shown that the completed FR-product cannot always be computed by the disjunctive FR-product of completed scenarios, even when the scenarios are self-complete. That is, we disprove Conjecture \#2 of Ref.~\cite{sainz2018multipartite}. In addition, we further show (by building on the results of Ref.~\cite[Appendix G]{dourdent2025paradox}) that already for two copies of a contextuality scenario one can see that $H {}^\disj \otimes H$ may yield hyperedges that do not correspond to valid process functions in general  \cite{kunjwal2023nonclassicality}. From these two results one can conclude that the isomorphism between the `valid process functions on $n$ devices' and the `disjunctive product of $n$ single-party Bell scenarios' breaks when the scenarios being composed are general ones. 
We see that, when one leaves the strict version of the device-independent paradigm and leverages further local information (on how local measurements relate) to explore nonclassicality of nature, then the hyperedges of the composite systems -- when obtained via the disjunctive FR-product -- no longer nicely relate to process functions on arbitrary devices. 
An interesting question is how the hyperedges of the completed FR-product of arbitrary scenarios may or not relate to quasi-process functions that do not yield paradoxes on that specific composite system. 

Our paper also disproves the Conjecture in Footnote \#10 of Ref.~\cite{sainz2018multipartite}: there is a self-complete scenario for which $H \overbar{\otimes} H \neq H {}^{\mathrm{FR}}\otimes H$. It is worth mentioning that this conjecture holds true for the specific case of composition of single-party Bell scenarios, since (i) the hierarchy of composition rules between the minimal FR-product and the maximal FR-product collapses when only two scenarios are composed \cite{sainz2018multipartite}, and (ii) for Bell scenarios, the disjunctive FR-product and the completed FR-product coincide as well as the maximal FR-product and the disjunctive FR-product \cite{leitherer2026bell}. In addition our results show  how the hierarchy of composition rules -- which collapses for bipartite Bell scenarios as we just stated -- no longer collapses for bipartite compositions of general scenarios, as sometimes $E(H \, {}^\mathrm{max} \otimes H) \subsetneq E(H \, {}^\disj \otimes H) $ and $E(H \, {}^\disj \otimes H) \subsetneq E(H \, \overbar{\otimes} H) $. 
This highlights how rich the simplest case of composition of two scenarios can be, and that interesting physics may emerge already in bipartite compositions. It is worth mentioning that Ref.~\cite[Sec.~3]{sainz2018multipartite} states that ${}^\mathrm{min} \otimes \equiv {}^\disj \otimes$ for bipartite composition of general scenarios (which we have here shown does not hold) but only applies it to Bell scenarios (and hence its results hold).

We have also presented a sufficient condition on a collection of hypergraphs for their disjunctive FR-product to be self-complete -- Lemma \ref{lem:sc}.  This sufficient condition is shown not to be necessary, and serves as a replacement for the now-disproven Conjecture \#2 of Ref.~\cite{sainz2018multipartite}. Whether there exists an alternative sufficient condition that is also necessary is left as an open question. In addition, Conjecture \#2 of Ref.~\cite{sainz2018multipartite} was further meant to provide a tool with which to compute the completed FR-product of a collection of scenarios. We showed that the conditions in Lemma \ref{lem:sc} are easy to certify (a linear program), and hence this provides a systematic way to compute the completed FR-product for these particular scenarios. It remains an open question how to systematically compute the completed FR-product for arbitrary contextuality scenarios.

\section*{Acknowledgments}

We thank John H.~Selby, Hippolyte Dourdent, and Andreas Leitherer for fruitful discussions during the initial stages of the project. We also thank Hippolyte Dourdent for pointing us to Ref.~\cite{dourdent2025paradox} at later stages in the project. 
We acknowledge the use of the AI assistant ChatGPT GPT-6 model (OpenAI).
This work is partially carried out under the IRA Programme, project no.~FENG.02.01-IP.05-0006/23, financed by the FENG program 2021-2027, Priority FENG.02, Measure FENG.02.01., with the support of the FNP.

\bibliographystyle{quantum}
\bibliography{biblio}

\appendix

\section{Examples from Section \ref{se:motex}}\label{ap:proofex}

Let us begin with the proof of Prop.~\ref{pr:ex18}, whose statement we include below. 

\bigskip

\noindent \textbf{Proposition \ref{pr:ex18}.}
Consider the contextuality scenarios $H_1$ and $H_2$, depicted in Fig.~\ref{fig:aqc3}, with vertex sets
\begin{align}
    V(H_1)=V(H_2)=\{A,B,C,a,b,c,\alpha,\beta,\gamma,q_1,q_2,q_3,q_4,w_0,w_1,w_2,w_3,w_4\}\,,  \label{eq:aver18}
\end{align}
and hyperedge sets
\begin{align}
    E(H_1) =  \{&\{A,B,c\}, \{a,B,C\},\{A,b,C\},\{a,\beta,c\},\{\alpha,b,c\},\{a,b,\gamma\}, \\ \nonumber &\{B,q_1,w_0\},\{q_1,q_2,w_1\}, \{q_2,q_3,w_2\},\{q_3,q_4,w_3\},\{B,q_4,w_4\} \}\,,\\
    E(H_2) =  \{&\{A,B,c\}, \{a,B,C\},\{A,b,C\},\{a,\beta,c\},\{\alpha,b,c\},\{a,b,\gamma\},\\ \nonumber 
    &\{B,q_1,w_0\},\{q_1,q_2,w_1\}, \{q_2,q_3,w_2\},\{q_3,q_4,w_3\},\{B,q_4,w_4\},\{\alpha,\beta,\gamma\} \} \,.
\end{align}
These scenarios satisfy the following: 
\begin{align} \label{eq:aneqor2}
    \mathrm{OrthoGraph}(H_1) &\neq \mathrm{OrthoGraph}(H_2)\,,\\
    \overbar{H}_1 &\neq \overbar{H}_2\,, \label{eq:aneqcom2}\\
    \mathcal{Q}_1(H_1) &= \mathcal{Q}_1(H_2)\,,\label{eq:aeqaq2} \\
    \mathcal{C}(H_i) \neq &\mathcal{Q}_1(H_i) \neq \mathcal{G}(H_i)\,, \quad i=1,2. \label{eq:ahiernocol}
\end{align}
\begin{proof}
To see that Eq.~\eqref{eq:aneqor2} holds, notice that the pair of vertices $\{\alpha,\beta\}$ forms an edge in $\mathrm{OrthoGraph}(H_2)$, but are not orthogonal in $H_1$, hence $\mathrm{OrthoGraph}(H_1) \neq \mathrm{OrthoGraph}(H_2)$. 

To see that Eq.~\eqref{eq:aneqcom2} holds, consider the behaviour $\mathbf{p} \in \mathcal{G}(H_1)$ given by 
\begin{align}
\mathbf{p} = \left[\frac{2}{5},\frac{2}{5},\frac{2}{5},\frac{1}{5},\frac{1}{5},\frac{1}{5},\frac{3}{5},\frac{3}{5},\frac{3}{5},\frac{2}{5},\frac{2}{5},\frac{2}{5},\frac{2}{5},\frac{1}{5},\frac{1}{5},\frac{1}{5},\frac{1}{5},\frac{1}{5}\right]\,,    
\end{align}
where the vertices are presented in the order of Eq.~\eqref{eq:aver18}. Since $p(\alpha)+p(\beta)+p(\gamma) = \frac{9}{5}>1$, this shows that $\{\alpha,\beta,\gamma\}$ does not form a virtual edge of $H_1$. The triple $\{\alpha,\beta,\gamma\}$, however, is a hyperedge in $H_2$ (and so in $\overbar{H}_2$), hence $\overbar{H}_1 \neq \overbar{H}_2$.

Let us now prove the first part of Eq.~\eqref{eq:ahiernocol}, i.e., $\mathcal{C}(H_i) \neq \mathcal{Q}_1(H_i)$ for $i=1,2$. For this, notice that\footnote{This follows by noticing that the subset of vertices $\{B,q_1,q_2,q_3,q_4\}$ forms the so-called KCBS scenario \cite{klyachko2008simple}.} 
\begin{align}
    p(B)+\sum_{k=1}^4 p(q_k) \leq 2\,. \label{eq:kcbs}
\end{align}
We will now present a quantum behaviour valid for both $H_1$ and $H_2$ and that violates Eq.~\eqref{eq:kcbs}. For this, we will leverage the KCBS quantum model \cite{klyachko2008simple}. Consider a qutrit system prepared in state $\ket{\psi} = [0,\, 0,\, 1]^\mathrm{T}$. Consider the vectors $\ket{v_k} = [\sin \theta \cos(4\pi k /5),\, \sin \theta \sin(4\pi k /5),\, \cos \theta]^\mathrm{T}$ for $k=0:4$ with $\cos^2\theta = 1/\sqrt{5}$. Now assign the following projectors to the following vertices: 
\begin{align}
    \Pi_{B} &= \ket{v_0} \bra{v_0} \,,\\
    \Pi_{q_k} &= \ket{v_k} \bra{v_k} \,, \quad k=1:4 \,\\
    \Pi_{w_k} &= \id - \ket{v_k} \bra{v_k}-\ket{v_{k+1}} \bra{v_{k+1}}\,,\, \quad k=0:4\,,\quad \ket{v_5} \equiv \ket{v_0}\,.
\end{align}
Choose two mutually orthogonal rank-one projectors $\Pi^{(1)}$ and $\Pi^{(2)}$ that satisfy $\bra{\psi} \Pi^{(k)} \ket{\psi} = (1-1/\sqrt{5})/2$ for $k=1,2$, are orthogonal to $\Pi_B$, and satisfy $\Pi^{(1)} + \Pi^{(2)} + \Pi_B = \id$\footnote{Choose an appropriate orthonormal basis of the two-dimensional subspace orthogonal to $\ket{v_0}$.}. Then define the following: 
\begin{align}
    \Pi_B = \Pi_b = \Pi_\beta\,, \\
    \Pi_A = \Pi_a = \Pi_\alpha = \Pi^{(1)}\,, \\
    \Pi_C = \Pi_c = \Pi_\gamma = \Pi^{(2)} \,.   
\end{align}
By definition, these projectors assigned to the vertices of $V(H_k)$ form a complete projective measurement for each hyperedge in $E(H_k)$ ($k=1,2$), and hence together with $\ket{\psi}$ define a quantum model $\mathbf{p}_Q$ that belongs to both $\mathcal{Q}(H_1)$ and $\mathcal{Q}(H_2)$. However,  $p_Q(B)+\sum_{k=1}^4 p_Q(q_k) = \sqrt{5} > 2$, hence $\mathbf{p}_Q \not\in \mathcal{C}(H_k)$ for $k=1,2$. Since quantum behaviours form a subset of almost-quantum ones, it follows that $\mathcal{C}(H_i) \neq \mathcal{Q}_1(H_i)$.  

Let us now prove the second part of Eq.~\eqref{eq:ahiernocol}, i.e., $\mathcal{Q}_1(H_i) \neq \mathcal{G}(H_i)$ for $i=1,2$. For this, consider the behaviour 
\begin{align}
    \mathbf{p}_G = \left[ \frac{1}{4},\, \frac{1}{2},\, \frac{1}{4},\, \frac{1}{4},\, \frac{1}{2},\, \frac{1}{4},\, \frac{1}{4},\, \frac{1}{2},\, \frac{1}{4},\, \frac{1}{2},\,\frac{1}{2},\,\frac{1}{2},\,\frac{1}{2},\,0,\,0,\,0,\,0,\,0 \right]\,. 
\end{align}
One can check that this behaviour belongs to $\mathcal{G}(H_k)$ but violates the consistent exclusivity principle already for two copies of the scenario -- i.e., $H_k \, {}^{\mathrm{FR}}\otimes H_k$  -- for $k=1,2$. Indeed, consider the five bipartite events
\begin{align}
    I = \{(B,B),\, (q_1,q_2),\, (q_2,q_4),\, (q_3,q_1),\,(q_4,q_3)\}\,.
\end{align}
Consistent exclusivity -- which behaviours in $\mathcal{Q}_1$ always satisfy -- imposes that $\sum_{v \in I} p(v) \leq 1$. However, the bipartite behaviour $\mathbf{p}^{(2)}_G = \mathbf{p}_G\times \mathbf{p}_G$ gives $\sum_{v \in I} p^{(2)}_G(v) = \frac{5}{4} >1 $. 
From this follows that $\mathcal{Q}_1(H_i) \neq \mathcal{G}(H_i)$ for $i=1,2$.

Now let us finally show Eq.~\eqref{eq:aeqaq2}. First, notice that $H_1$ is a subhypergraph of $H_2$, and hence $\mathcal{Q}_1(H_2) \subseteq \mathcal{Q}_1(H_1)$. Now we will show that any almost-quantum behaviour in $H_1$ is also an almost-quantum behaviour in $H_2$. For this, notice that the only extra constraints that an almost-quantum model of $H_1$ needs to satisfy for this to be the case are the ones that arise from: 
\begin{compactitem}
\item the fact that $\{\alpha,\beta,\gamma\}$ is a hyperedge in $H_2$, and
\item the extra orthogonality constraints that $H_2$ imposes: $\alpha \perp \beta$, $\alpha \perp \gamma$, and $\beta \perp \gamma$. 
\end{compactitem}
So let us consider a (fixed but arbitrary) almost-quantum model $\mathbf{p} \in \mathcal{Q}_1(H_1)$, and represent it using a unit vector $\ket{\Psi}$ and a set of unit vectors $\{\ket{v}\}_{v\in V(H_1)}$ as per Def.~\ref{def:aqaflsc}. These satisfy: 
\begin{align}
p(v) &= \tr{\ket{v}\bra{v} \ket{\Psi} \bra{\Psi}}{} \quad \forall v \in V(H_1) \,,\\
u &\perp_{H_1} v \quad \Rightarrow \langle u \vert v \rangle = 0\,.
\end{align}
We want to show that this choice of vectors $\{\ket{v}\}$ and vector $\ket{\Psi}$ also define an almost-quantum behaviour in $H_2$. First notice that the consistent exclusivity principle \cite{acin2015combinatorial} implies both $p(A)+p(B)+p(C) \leq 1$ and $p(a)+p(b)+p(c) \leq 1$. Moreover, the normalisation constraints imposed by the hyperedges of $H_1$ imply that those two inequalities are always saturated. All these linear constraints together imply that $p(\alpha)+p(\beta)+p(\gamma)=1$ for any behaviour in $\mathcal{Q}_1(H_1)$. Hence, $\mathbf{p} \in \mathcal{G}(H_2)$. 
Now what we also must show is that $\langle \alpha \vert \beta \rangle = 0$, $\langle \alpha \vert \gamma \rangle = 0$, and $\langle \beta \vert \gamma \rangle = 0$. 
Now, Ref.~\cite[Prop.~6.3.1]{acin2015combinatorial} implies that  $\sum_{v\in e} \ket{v} = \ket{\Psi}$ for all $e \in E(H_1)$. 
In addition, since $\{a,b,c\}$ are pairwise orthogonal in $H_1$ and their probabilities are normalised, this implies that $\sum_{v\in \{a,b,c\}} \ket{v} = \ket{\Psi}$. 
From this follows that $\ket{a}=\ket{\alpha}$, $\ket{b}=\ket{\beta}$, and $\ket{c}=\ket{\gamma}$. 
The orthogonality relations among $\{\ket{a},\ket{b},\ket{c}\}$ then imply the required orthogonality relations among $\{\ket{\alpha},\ket{\beta},\ket{\gamma}\}$. 
Therefore, any almost-quantum behaviour in $H_1$ automatically satisfies the constraints for it to be an almost-quantum behaviour in $H_2$, which shows that $\mathcal{Q}_1(H_1) \subseteq \mathcal{Q}_1(H_2)$. It follows that $\mathcal{Q}_1(H_1) = \mathcal{Q}_1(H_2)$, which concludes the proof. 
\end{proof}

\bigskip

Next, let us present a couple of examples that also show that equal orthogonality graphs is \textit{not necessary} for the almost-quantum sets of two scenarios to coincide. In both cases we do not demand that the sets of classical, almost-quantum, and general behaviours are different, and in one case we further allow the scenarios to have the same completion. 

\bigskip

\begin{figure}
\centering
\subfigure[Scenario $H_1$]{
\begin{tikzpicture}[scale=1]

	\node[draw,shape=circle,fill,scale=.5] (C) at (-30:1.9) {} ;
	\node[draw,shape=circle,fill,scale=.5] (B) at (90:1.9) {} ;
    \node[draw,shape=circle,fill,scale=.5]  (A) at (210:1.9) {} ;
    \node[draw,shape=circle,fill,scale=.5]  (c) at ($(A)!0.5!(B)$) {} ;
    \node[draw,shape=circle,fill,scale=.5]  (b) at ($(A)!0.5!(C)$) {} ;
    \node[draw,shape=circle,fill,scale=.5]  (a) at ($(C)!0.5!(B)$) {} ;
    \node[draw,shape=circle,fill,scale=.5]  (alpha) at ($(b)!0.5!(c)$) {} ;
    \node[draw,shape=circle,fill,scale=.5]  (beta) at ($(a)!0.5!(c)$) {} ;
    \node[draw,shape=circle,fill,scale=.5]  (gamma) at ($(b)!0.5!(a)$) {} ;

    \draw[thick,gray, rotate around={60:(c)}] (c) ellipse (2cm and .3cm) ;
    \draw[thick,gray, rotate around={120:(a)}] (a) ellipse (2cm and .3cm) ;
    \draw[thick,gray, rotate around={0:(b)}] (b) ellipse (2cm and .3cm) ;

    \draw[thick,gray, rotate around={60:(gamma)}] (gamma) ellipse (1.3cm and .2cm) ;
    \draw[thick,gray, rotate around={120:(alpha)}] (alpha) ellipse (1.3cm and .2cm) ;
    \draw[thick,gray, rotate around={0:(beta)}] (beta) ellipse (1.3cm and .2cm) ;


    \node at (-30:2.5) {$C$};
    \node at (90:2.5) {$B$};
    \node at (210:2.5) {$A$};

    \node at (30:1.5) {$a$};
    \node at (150:1.5) {$c$};
    \node at (270:1.5) {$b$};

    \node at (-30:0.8) {$\gamma$};
    \node at (90:0.8) {$\beta$};
    \node at (210:0.8) {$\alpha$};

\end{tikzpicture}
}    
\subfigure[Scenario $H_2$]{
\begin{tikzpicture}[scale=1]

\node[draw,shape=circle,fill,scale=.5] (C) at (-30:1.9) {} ;
	\node[draw,shape=circle,fill,scale=.5] (B) at (90:1.9) {} ;
    \node[draw,shape=circle,fill,scale=.5]  (A) at (210:1.9) {} ;
    \node[draw,shape=circle,fill,scale=.5]  (c) at ($(A)!0.5!(B)$) {} ;
    \node[draw,shape=circle,fill,scale=.5]  (b) at ($(A)!0.5!(C)$) {} ;
    \node[draw,shape=circle,fill,scale=.5]  (a) at ($(C)!0.5!(B)$) {} ;
    \node[draw,shape=circle,fill,scale=.5]  (alpha) at ($(b)!0.5!(c)$) {} ;
    \node[draw,shape=circle,fill,scale=.5]  (beta) at ($(a)!0.5!(c)$) {} ;
    \node[draw,shape=circle,fill,scale=.5]  (gamma) at ($(b)!0.5!(a)$) {} ;

    \draw[thick,gray, rotate around={60:(c)}] (c) ellipse (2cm and .3cm) ;
    \draw[thick,gray, rotate around={120:(a)}] (a) ellipse (2cm and .3cm) ;
    \draw[thick,gray, rotate around={0:(b)}] (b) ellipse (2cm and .3cm) ;

    \draw[thick,gray, rotate around={60:(gamma)}] (gamma) ellipse (1.3cm and .2cm) ;
    \draw[thick,gray, rotate around={120:(alpha)}] (alpha) ellipse (1.3cm and .2cm) ;
    \draw[thick,gray, rotate around={0:(beta)}] (beta) ellipse (1.3cm and .2cm) ;

\draw[thick,darkgreen] plot [smooth cycle,tension=.8] coordinates { (-30:0.6) (90:0.6) (210:0.6) } ;

    \node at (-30:2.5) {$C$};
    \node at (90:2.5) {$B$};
    \node at (210:2.5) {$A$};

    \node at (30:1.5) {$a$};
    \node at (150:1.5) {$c$};
    \node at (270:1.5) {$b$};

    \node at (-30:0.8) {$\gamma$};
    \node at (90:0.8) {$\beta$};
    \node at (210:0.8) {$\alpha$};

\end{tikzpicture}
}      
\caption{Scenarios $H_1$ and $H_2$ used in Prop.~\ref{pr:ex9} to provide a counterexample to Conjecture \#1 of Ref.~\cite{sainz2018multipartite} where the scenarios have different completions. Notice that $H_2$ has an extra hyperedge ($\{\alpha,\beta,\gamma\}$, depicted in green) but is otherwise identical to $H_1$.}
\label{fig:aqc2}
\end{figure}
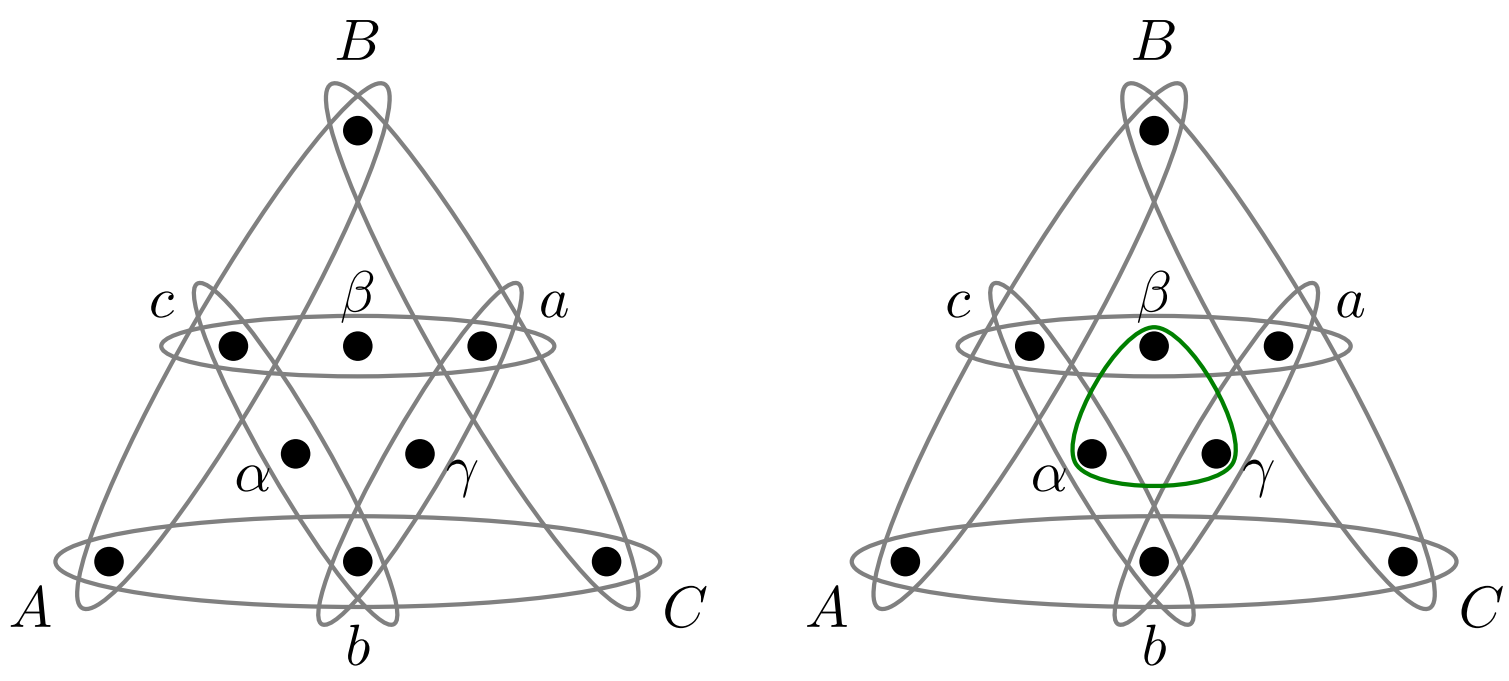

\begin{prop}\label{pr:ex9}
Consider the contextuality scenarios $H_1$ and $H_2$, depicted in Fig.~\ref{fig:aqc2}, with vertex sets
\begin{align}
    V(H_1)=V(H_2)=\{A,B,C,a,b,c,\alpha,\beta,\gamma\}\,, \label{eq:ver9}
\end{align}
and hyperedge sets
\begin{align}
    E(H_1) &= \{\{A,B,c\}, \{a,B,C\},\{A,b,C\},\{a,\beta,c\},\{\alpha,b,c\},\{a,b,\gamma\}\}\,,\\
    E(H_2) &= \{\{A,B,c\}, \{a,B,C\},\{A,b,C\},\{a,\beta,c\},\{\alpha,b,c\},\{a,b,\gamma\},\{\alpha,\beta,\gamma\}\}\,.
\end{align}
These scenarios satisfy the following: 
\begin{align} \label{eq:neqor}
    \mathrm{OrthoGraph}(H_1) &\neq \mathrm{OrthoGraph}(H_2)\,,\\
    \overbar{H}_1 &\neq \overbar{H}_2\,, \label{eq:neqcom}\\
    \mathcal{Q}_1(H_1) &= \mathcal{Q}_1(H_2)\,. \label{eq:eqaq}
\end{align}
\end{prop}
\begin{proof} \quad \\
To see that Eq.~\eqref{eq:neqor} holds, notice that the pair of vertices $\{\alpha,\beta\}$ belongs to a hyperedge in $H_2$ and hence forms an edge in $\mathrm{OrthoGraph}(H_2)$, while they are not orthogonal in $H_1$ (don't share a hyperedge there) and hence do not form an edge in $\mathrm{OrthoGraph}(H_1)$. Hence, $\mathrm{OrthoGraph}(H_1) \neq \mathrm{OrthoGraph}(H_2)$.
To see that Eq.~\eqref{eq:neqcom} holds, consider the behaviour $\mathbf{p} \in \mathcal{G}(H_1)$ given by ${\mathbf{p} = [\frac{2}{5},\frac{2}{5},\frac{2}{5},\frac{1}{5},\frac{1}{5},\frac{1}{5},\frac{3}{5},\frac{3}{5},\frac{3}{5}]}$, where the vertices are presented in the order of Eq.~\eqref{eq:ver9}. Since $p(\alpha)+p(\beta)+p(\gamma) = \frac{9}{5}>1$, this shows that $\{\alpha,\beta,\gamma\}$ does not form a virtual edge of $H_1$. The triple $\{\alpha,\beta,\gamma\}$, however, is a hyperedge in $H_2$, hence $\overbar{H}_1 \neq \overbar{H}_2$.

Now we need to show that Eq.~\eqref{eq:eqaq} holds, and we will do it in three steps. \\
\noindent $\bullet$ We first show that $\mathcal{Q}_1(H_1) \subseteq \mathcal{G}(H_2)$. Let $\mathbf{p}$ be a behaviour in $\mathcal{Q}_1(H_1)$. Because of the consistent exclusivity principle \cite{acin2015combinatorial}, both $p(A)+p(B)+p(C) \leq 1$ and $p(a)+p(b)+p(c) \leq 1$. Moreover, the normalisation constraints imposed by the hyperedges of $H_1$ imply that those two inequalities are always saturated. All these linear constraints together imply that $p(\alpha)+p(\beta)+p(\gamma)=1$. This means that $\mathbf{p}$ satisfies all the normalisation constraints imposed by the hyperedges of $H_2$. Therefore, $\mathbf{p} \in \mathcal{G}(H_2) \,\, \forall \,\, \mathbf{p} \in \mathcal{Q}_1(H_1) $, which implies $\mathcal{Q}_1(H_1) \subseteq \mathcal{G}(H_2)$. \\
\noindent $\bullet$ Notice that the conditions for a behaviour $\mathbf{p}$ to be in $\mathcal{Q}_1(H_2)$ include all the conditions for $\mathbf{p} \in \mathcal{Q}_1(H_1)$, since $\mathrm{OrthoGraph}(H_1)$ is a subgraph of $\mathrm{OrthoGraph}(H_2)$  and $\mathcal{Q}_1(H_1) \subseteq \mathcal{G}(H_2)$. Hence, $\mathcal{Q}_1(H_2) \subseteq \mathcal{Q}_1(H_1)$. \\
\noindent $\bullet$ Third, let us show that  $\mathcal{C}(H_2) = \mathcal{Q}_1(H_1)$. The constraints $p(A)+p(B)+p(C) = 1$ and $p(a)+p(b)+p(c) = 1$ that we previously showed, together with the normalisation constraints imposed by the hyperedges of $H_1$ imply that: 
\begin{align}
    &p(A)=p(a)\,,\quad p(B)=p(b)\,,\quad p(C)=p(c)\,,\\
    &p(\alpha)=p(a)\,,\quad p(\beta)=p(b)\,,\quad  p(\gamma)=p(c)\,. \nonumber
\end{align}
Hence, every behaviour $\mathbf{p} \in \mathcal{Q}_1(H_1)$ must be parametrised as
\begin{align}
    \mathbf{p} = [r,\,  s,\, t,\, r,\,  s,\, t,\, r,\,  s,\, t]\,, \quad r,s,t \geq 0\,,\quad r+s+t=1\,.
\end{align}
Then notice that the following three behaviours are deterministic and valid in $\mathcal{G}(H_2)$, and hence belong to $\mathcal{C}(H_2)$:
\begin{align}
    &\mathbf{p}_{C1} = [1,\, 0 ,\, 0 ,\, 1,\, 0 ,\, 0 ,\, 1,\, 0 ,\, 0]\, ,\\
    &\mathbf{p}_{C2} = [0,\, 1 ,\, 0 ,\, 0,\, 1 ,\, 0 ,\, 0,\, 1 ,\, 0]\, ,\\
    &\mathbf{p}_{C3} = [0,\, 0 ,\, 1 ,\, 0,\, 0 ,\, 1 ,\, 0,\, 0 ,\, 1]\,.
\end{align}
Therefore, $\mathbf{p} \in \mathcal{Q}_1(H_1)$ only if $\mathbf{p} = r \, \mathbf{p}_{C1} + s \, \mathbf{p}_{C2} + t \, \mathbf{p}_{C3} \,, \quad r,s,t \geq 0\,,\quad r+s+t=1$, which implies $\mathbf{p} \in \mathcal{C}(H_2)$. This shows that $\mathcal{Q}_1(H_1) \subseteq \mathcal{C}(H_2) \subseteq \mathcal{Q}_1(H_2) \subseteq \mathcal{Q}_1(H_1)$, which shows that $\mathcal{Q}_1(H_1) = \mathcal{C}(H_2)$ and consequently $\mathcal{Q}_1(H_2) = \mathcal{Q}_1(H_1)$. This concludes the proof. 
\end{proof}

\bigskip

\begin{figure}
\centering
\subfigure[Scenario $H_1$]{
\begin{tikzpicture}[scale=1]

	\node[draw,shape=circle,fill,scale=.5] at (0,0.2) {} ;
	\node[draw,shape=circle,fill,scale=.5] at (2.5,0.2) {} ;
	\draw[thick,black] (1.25,0.2) ellipse (1.7cm and .4cm) ;
	
	\node[draw,shape=circle,fill,scale=.5] at (0,-1.9) {} ;
	\node[draw,shape=circle,fill,scale=.5] at (2.5,-1.9) {} ;
    \node[draw, color=white] at (-1,-2.5) {};
    \node[draw, color=white] at (4,0) {};

\draw[thick,black, rotate around={90:(0,-0.85)}] (0,-0.85) ellipse (1.5cm and .4cm) ;
\draw[thick,black, rotate around={90:(2.5,-0.85)}] (2.5,-0.85) ellipse (1.5cm and .4cm) ;

\node at (-0.5,0.5) {$v_2$};
\node at (-0.5,-2.4) {$v_1$};
\node at (3,0.5) {$v_3$};
\node at (3, -2.4) {$v_4$};
\end{tikzpicture}
}    
\subfigure[Scenario $H_2$]{
\begin{tikzpicture}[scale=1]

	\node[draw,shape=circle,fill,scale=.5] at (0,0.2) {} ;
	\node[draw,shape=circle,fill,scale=.5] at (2.5,0.2) {} ;
	\draw[thick,black] (1.25,0.2) ellipse (1.7cm and .4cm) ;
	
	\node[draw,shape=circle,fill,scale=.5] at (0,-1.9) {} ;
	\node[draw,shape=circle,fill,scale=.5] at (2.5,-1.9) {} ;
	\draw[thick,black] (1.25,-1.9) ellipse (1.7cm and .4cm) ;
    \node[draw, color=white] at (-1,-2.5) {};
    \node[draw, color=white] at (4,0) {};

\draw[thick,black, rotate around={90:(0,-0.85)}] (0,-0.85) ellipse (1.5cm and .4cm) ;
\draw[thick,black, rotate around={90:(2.5,-0.85)}] (2.5,-0.85) ellipse (1.5cm and .4cm) ;

\node at (-0.5,0.5) {$v_2$};
\node at (-0.5,-2.4) {$v_1$};
\node at (3,0.5) {$v_3$};
\node at (3, -2.4) {$v_4$};

\end{tikzpicture}
}      
\caption{Scenarios $H_1$ and $H_2$ used in Prop.~\ref{pr:aqc} to provide a counterexample to Conjecture \#1 of Ref.~\cite{sainz2018multipartite}.}
\label{fig:aqc}
\end{figure}
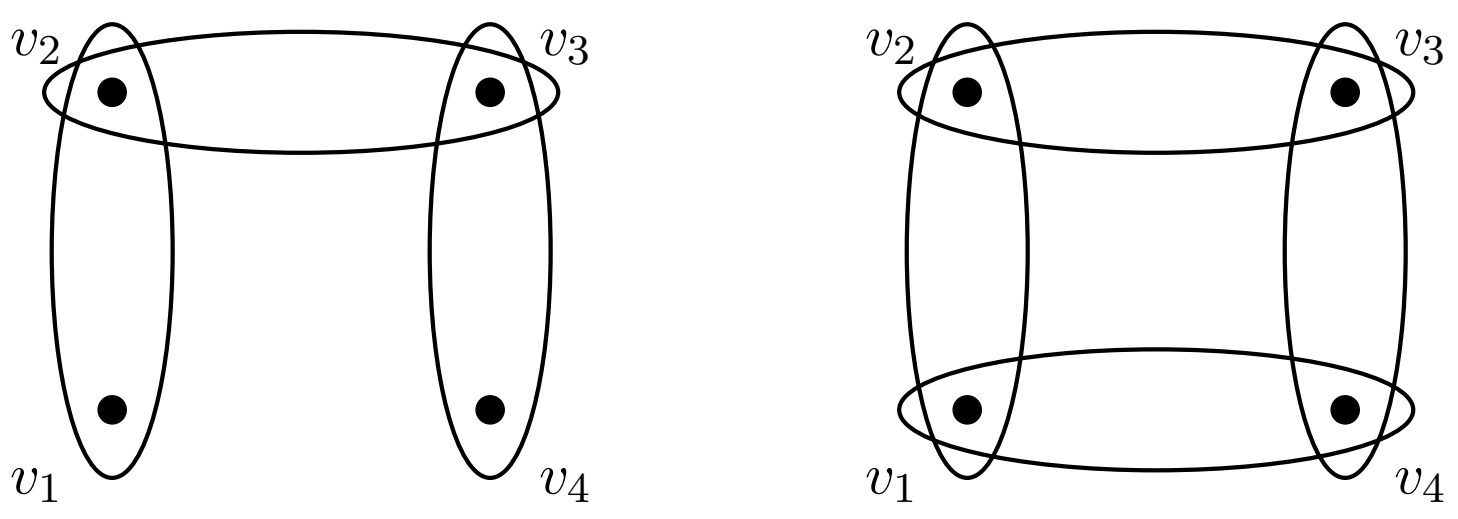

\begin{prop}\label{pr:aqc}
Consider the contextuality scenarios $H_1$ and $H_2$, depicted in Fig.~\ref{fig:aqc}, with vertex sets
\begin{align}
    V(H_1) = V(H_2) = \{v_1,v_2,v_3,v_4\}\,,
\end{align}
and hyperedge sets
\begin{align}
    E(H_1) &= \{\{v_1,v_2\} , \{v_2,v_3\} , \{v_3,v_4\} \}\,,\\
    E(H_2) &= \{\{v_1,v_2\} , \{v_2,v_3\} , \{v_3,v_4\} , \{v_4,v_1\} \}\,.
\end{align}
These scenarios satisfy the following: 
\begin{align} \label{eq:neqori}
    \mathrm{OrthoGraph}(H_1) &\neq \mathrm{OrthoGraph}(H_2)\,,\\
    \mathcal{Q}_1(H_1) &= \mathcal{Q}_1(H_2)\,. \label{eq:eqaqii}
\end{align}
\end{prop}
\begin{proof}
First notice that the  orthogonality graph of $H_1$ has edge set $E(\mathrm{OrthoGraph}(H_1)) = E(H_1)$ and the orthogonality graph of $H_2$ has edge set $E(\mathrm{OrthoGraph}(H_2)) = E(H_2)$. Hence, $\mathrm{OrthoGraph}(H_2)$ has an extra edge which $\mathrm{OrthoGraph}(H_1)$ doesn't feature -- $\{v_4,v_1\}$ -- which implies Eq.~\eqref{eq:neqori}.

Now we need to see that the two scenarios admit the same set of almost-quantum behaviours. For this, first notice that the normalisation constraints given by the hyperedges in $H_1$ imply the normalisation constraints given by the hyperedges in $H_2$. Hence, $\mathcal{G}(H_1)=\mathcal{G}(H_2)$. Next, notice that the normalisation constraints further imply that an arbitrary behaviour in $\mathcal{G}(H_1)$ is parametrised by $\mathbf{p}_{H_1}(t) = [t,1-t, t, 1-t]$, with $t \in [0,1]$. This means that any behaviour in $\mathcal{G}(H_1)$ can be decomposed as a convex combination of the two deterministic behaviours $\mathbf{p}_{H_1}(0)$ and $\mathbf{p}_{H_1}(1)$: 
\begin{align}
    \mathbf{p}_{H_1} \in \mathcal{G}(H_1) \quad \mathrm{iff} \quad \exists \,  t \in [0,1] \quad \mathrm{st} \quad \mathbf{p}_{H_1} = t \, \mathbf{p}_{H_1}(1) + (1-t) \, \mathbf{p}_{H_1}(0) \,.
\end{align}
From this follows that the set of allowed behaviours in $H_1$ (and hence in $H_2$) coincides with the set of classical behaviours in $H_1$, denoted by $\mathcal{C}(H_1)$. Since $\mathcal{C}(H_1) \subseteq \mathcal{Q}_1(H_1) \subseteq \mathcal{G}(H_1)$, it follows that this hierarchy collapses: $\mathcal{C}(H_1) = \mathcal{Q}_1(H_1) = \mathcal{G}(H_1)$. Similarly, $\mathcal{C}(H_2) = \mathcal{Q}_1(H_2) = \mathcal{G}(H_2)$. Now, since $\mathcal{G}(H_1)=\mathcal{G}(H_2)$, this implies Eq.~\eqref{eq:eqaqii}, which concludes the proof. 
\end{proof}

\section{Proofs from Section \ref{se:counter}}\label{ap:counter}

We begin with the proof of Prop.~\ref{pr:no2}, whose statement we include below. 

\bigskip

\noindent \textbf{Proposition \ref{pr:no2}. }
The contextuality scenario $H=(V,E)$, depicted in Fig.~\ref{fig:pentagon}, with vertex set
\begin{align}
    V = \{v_1,v_2,v_3,v_4,v_5\}\,,
\end{align}
and hyperedge set
\begin{align}
    E = \{\{v_1,v_2,v_3\} , \{v_3,v_4,v_5\} , \{v_1,v_5\} \}
\end{align}
satisfies
\begin{align} \label{eq:Hsc}
    \overbar{H} &= H\,,\\ \label{eq:Hpdif}
    \overbar{H} \,  {}^{\disj}\otimes \overbar{H} & \neq \overbar{H} \,  \overbar{\otimes} \,  \overbar{H}\,.
\end{align}
\begin{proof}
    Let us denote $e_1 = \{v_1,v_2,v_3\}$, $e_2 = \{v_3,v_4,v_5\}$, and $ e_3 =  \{v_1,v_5\}$. We start by showing $H$ satisfies Eq.~\eqref{eq:Hsc}, and then show that Eq.~\eqref{eq:Hpdif} holds. 

\bigskip

\noindent $\bullet \, H$ \textbf{is self-complete -- Eq.~\eqref{eq:Hsc} holds:} \\
We need to show that $H$ has no virtual edges. That is, the possible normalisation conditions satisfied by an arbitrary behaviour $\mathbf{p}_H \in \mathcal{G}(H)$ do not include any subset of vertices of $H$ beyond the hyperedges $E$. 

The normalisation conditions given by the hyperedges correspond to:
\begin{align}
    \begin{cases}
        p(v_1)+p(v_2)+p(v_3) &= 1 \,, \\
        p(v_3)+p(v_4)+p(v_5) &= 1 \,, \\
        p(v_1)+p(v_5) &= 1 \,.
    \end{cases}
\end{align}
If we denote $\alpha := p(v_1)$ and $\beta := p(v_3)$, we can parametrise an arbitrary $\mathbf{p}_H$ by
\begin{align}\label{eq:82}
    \mathbf{p}_H = [p(v_1),p(v_2),p(v_3),p(v_4),p(v_5)] = [\alpha,1-\alpha-\beta,\beta,\alpha-\beta,1-\alpha]\,,
\end{align}
which is a valid behaviour for $\alpha \in [0,1]$ and $\beta \in [0,\min\{\alpha,1-\alpha\}]$\,.

Consider the behaviour $\mathbf{p}$ given by $\alpha = \frac{2}{5}$ and $\beta = \frac{1}{10}$. 
Notice that the vertex set $V$ has 31 possible subsets $s$ (excluding  $\emptyset$). 
If we compute $\sum_{v \in s} p(v)$ for each of those 31 choices for $s$, we find that only three choices give $\sum_{v \in s} p(v) =1$, which are $s=e_1$, $s=e_2$, and $s=e_3$. Since no other subset of vertices have probabilities sum up to $1$ for $\mathbf{p}$ beyond the hyperedges themselves, this shows that $H$ has no virtual edges and hence is self-complete. 

\bigskip

\noindent $\bullet$ \textbf{Eq.~\eqref{eq:Hpdif} holds:} \\
To show this we need to find a hyperedge in $H \overbar{\otimes} H$ that is not a hyperedge in $H \, {}^{\disj}\otimes H$. Our candidate for such a hyperedge is the subset $S \subset V \times V$ given by 
\begin{align}
S = \{(v_2,v_2), (v_4,v_4),(v_4,v_3), (v_3,v_4),(v_3,v_5),(v_1,v_5),(v_5,v_3),(v_5,v_1)\}\,.
\end{align}
It is straightforward to see that $S \not\in E(H \, {}^{\disj}\otimes H)$, since the composite events $(v_2,v_2)$ and $(v_4,v_4)$ are not locally orthogonal. 

We will now see that $S$ is a virtual edge, and hence $S \in E(H \overbar{\otimes} H)$. For this, we need to see that 
\begin{align}\label{eq:normS}
    \sum_{v \in S} p(v) = 1
\end{align}
is valid for any joint behaviour $\mathbf{p}_{H \, {}^{\max}\otimes H} \in \mathcal{G}(H \, {}^{\max}\otimes H)$. We will do so by showing that Eq.~\eqref{eq:normS} follows from the normalisation constraints imposed by the hyperedges of $H \, {}^{\max}\otimes H$. 

First, recall that $E(H \, {}^{\comm}\otimes H) \subseteq E(H \, {}^{\max}\otimes H)$, hence hyperedges that arise from measurement protocols are valid hyperedges of $H \, {}^{\max}\otimes H$. We will be interested in 6 such hyperedges: 
\begin{compactitem}
\item $E_1$: is the correlated measurement in $H \, {}^{\max}\otimes H$ where we measure $e_1$ in the left-hand-side (lhs) scenario $H$, and then choose the measurement in the right-hand-side (rhs) scenario $H$ as follows: if outcome is $v_1$ choose $e_3$, if outcome is $v_2$ choose $e_1$, and if outcome is $v_3$ choose $e_3$. The vertices in this hyperedge are: 
\begin{align}
    E_1 = \{ (v_1,v_1),(v_1,v_5), 
    (v_2,v_1), (v_2,v_2),(v_2,v_3),
    (v_3,v_1),(v_3,v_5)\}\,.
\end{align}

\item $E_2$: is the correlated measurement where we measure $e_2$ in the lhs $H$, and then choose the measurement in the rhs $H$ as: if $v_3$ choose $e_2$, if $v_4$ choose $e_2$, and if $v_5$ choose $e_1$. The vertices in this hyperedge are: 
\begin{align}
    E_2 = \{ (v_3,v_3),(v_3,v_4),(v_3,v_5),
    (v_4,v_3),(v_4,v_4),(v_4,v_5),
    (v_5,v_1),(v_5,v_2),(v_5,v_3)\}\,.
\end{align}

\item $E_3$: is the correlated measurement where we measure $e_2$ in the lhs $H$, and then choose the measurement in the rhs $H$ as: if $v_3$ choose $e_1$, if $v_4$ choose $e_3$, and if $v_5$ choose $e_2$. The vertices in this hyperedge are: 
\begin{align}
    E_3 = \{ (v_3,v_1),(v_3,v_2),(v_3,v_3),
    (v_4,v_1),(v_4,v_5),
    (v_5,v_3),(v_5,v_4),(v_5,v_5)\}\,.
\end{align}

\item $E_4$: is the correlated measurement where we measure $e_1$ in the rhs $H$, and then choose the measurement in the lhs $H$ as: if $v_1$ choose $e_2$, if $v_2$ choose $e_1$, and if $v_3$ choose $e_3$. The vertices in this hyperedge are: 
\begin{align}
    E_4 = \{ (v_3,v_1),(v_4,v_1),(v_5,v_1),
    (v_1,v_2),(v_2,v_2),(v_3,v_2),
    (v_1,v_3),(v_5,v_3)\}\,.
\end{align}

\item $E_5$: is the correlated measurement where we measure $e_1$ in the rhs $H$, and then choose the measurement in the lhs $H$ as: if $v_1$ choose $e_1$, if $v_2$ choose $e_3$, and if $v_3$ choose $e_1$. The vertices in this hyperedge are: 
\begin{align}
    E_5 = \{ (v_1,v_1),(v_2,v_1),(v_3,v_1),
    (v_1,v_2),(v_5,v_2),
    (v_1,v_3),(v_2,v_3),(v_3,v_3)\}\,.
\end{align}

\item $E_6$: is the correlated measurement where we measure $e_2$ in the rhs $H$, and then choose the measurement in the lhs $H$ as: if $v_3$ choose $e_2$, if $v_4$ choose $e_2$, and if $v_5$ choose $e_3$. The vertices in this hyperedge are: 
\begin{align}
    E_6 = \{ (v_3,v_3),(v_4,v_3),(v_5,v_3),
    (v_3,v_4),(v_4,v_4),(v_5,v_4),
    (v_1,v_5),(v_5,v_5)\}\,.
\end{align}
\end{compactitem}
Since these are hyperedges of $H \, {}^{\max}\otimes H$ and $H \, {}^{\disj}\otimes H$ it follows that $\sum_{v \in E_k} p(v) = 1$ for all joint behaviour $\mathbf{p}_{H \, {}^{\disj}\otimes H} \in \mathcal{G}(H \, {}^{\disj}\otimes H)$, and for $k=1:6$.

Now, one needs to notice the following. On the one hand, 
\begin{align}
    \sum_{v \in E_1} p(v) + \sum_{v \in E_2} p(v) - \sum_{v \in E_3} p(v) + \sum_{v \in E_4} p(v) - \sum_{v \in E_5} p(v) + \sum_{v \in E_6} p(v) = 1+1-1+1-1+1 = 2\,,
\end{align}
and on the other hand
\begin{align}
    \sum_{v \in E_1} p(v) + \sum_{v \in E_2} p(v) - \sum_{v \in E_3} p(v) + \sum_{v \in E_4} p(v) - \sum_{v \in E_5} p(v) + \sum_{v \in E_6} p(v) = 2 \sum_{v \in S} p(v) \,.
\end{align}
Combining these two equations shows that Eq.~\eqref{eq:normS} is valid for any joint behaviour $\mathbf{p}_{H \, {}^{\max}\otimes H} \in \mathcal{G}(H \, {}^{\max}\otimes H)$, and hence $S$ corresponds to a virtual edge. 
\end{proof}

\bigskip

Next we present the proof of Lemma \ref{lem:sc}, whose statement we include below.

\bigskip

\noindent \textbf{Lemma \ref{lem:sc}.} 
Consider a set of contextuality scenarios $\{H_i\}$, such that each satisfies the following property: for each pair of vertices $u,v \in V(H_i)$ such that $u \not\perp v$ (including $u=v$) there exists a behaviour $\mathbf{p} \in \mathcal{G}(H_i)$ such that $p(u)=1$ and $p(v)>0$. Then, the disjunctive product of these scenarios is self-complete: 
\begin{align}
    {}^\disj \otimes_i H_i = \overbar{{}^\disj \otimes_i H_i} =: \overbar{\otimes}_i H_i. 
\end{align}
\begin{proof}

Consider two different vertices $\mathbf{u}$ and    $\mathbf{v}$ in the composite scenario $H = {}^\disj \otimes_i H_i$. If these two vertices are not locally-orthogonal in $H$, then each pair $(u_i,v_i)$ is not orthogonal in $H_i$. Consider then the behaviours $\mathbf{p}_i \in \mathcal{G}(H_i)$ such that $p_i(u_i) = 1$ and $p_i(v_i) >0$ for all $i$, which exist by the Lemma's assumptions. Hence, the product behaviour $\mathbf{p} \in \mathcal{G}(H)$ given by $p(\mathbf{x}) = \prod_i p_i(x_i)$ has the following properties: $p(\mathbf{u}) = 1$ and $p(\mathbf{v}) > 0$. Hence, $\mathbf{u}$ and    $\mathbf{v}$ cannot belong to a virtual hyperedge of $H$ (they already do not belong to $E(H)$ because they are not locally orthogonal). Therefore, any potential virtual hyperedge of $H$ must consist of sets of mutually-orthogonal vertices of $H$. 

What we need to do now is show that any such potential virtual edge already exists as an edge in $H$. That is, we need to show that if there exists a subset $s \subset V(H)$ of mutually-orthogonal events such that $\sum_{\mathbf{x} \in s} p(\mathbf{x}) = 1$ for every behaviour in $\mathcal{G}(H)$, then the constraint $\sum_{\mathbf{x} \in s} p(\mathbf{x}) = 1$ follows from a linear combination of the normalisation hyperedges and the No Signalling equalities. 

For this, notice first that there exists a behaviour in $\mathcal{G}(H)$ that has full support -- i.e., that it assigns non-zero values to each vertex. To construct this behaviour, recall that by assumption for each $H_i$ and for each $v \in V(H_i)$ there exists a behaviour $\mathbf{p}_{i,v} \in \mathcal{G}(H_i)$ such that $p_{i,v}(v) = 1$. By convexity, the behaviour $\mathbf{q}_i = \frac{1}{|V(H_i)|} \sum_{v\in V(H_i)} \mathbf{p}_{i,v}$ is then allowed in $\mathcal{G}(H_i)$ and assigns non-zero values to all the vertices. Then, define in $H$ the product behaviour $\mathbf{q}:= \prod_i \mathbf{q}_i$. It follows that $\mathbf{q}$ is an allowed behaviour in $\mathcal{G}(H)$ and assigns non-zero values to all the vertices $V(H)$.

We will now see that the existence of $\mathbf{q}$ implies that for any subset $s \subset V(H)$ of mutually-orthogonal events such that $\sum_{\mathbf{x} \in s} p(\mathbf{x}) = 1$ the rank condition 9b of Ref.~\cite{sainz2018multipartite} is satisfied. Following the notation of Ref.~\cite{sainz2018multipartite}, denote by $M_{\mathrm{Normalisation}}$ the matrix with $|V(H)|$ columns, and whose rows correspond to the constraints on behaviours in $\mathcal{G}(H)$ that arise from normalisation conditions at each $H_i$. Similarly, denote by $M_{\mathrm{NoSignalling}}$ the matrix with $|V(H)|$ columns, and whose rows correspond to the constraints on behaviours in $\mathcal{G}(H)$ that arise from No Signalling considerations among the $H_i$'s. Denote now by $M$ the vertical concatenation of these two matrices: $M_{\mathrm{Normalisation}} \, M_{\mathrm{NoSignalling}}$. Hence, a behaviour $\mathbf{p}$ belongs to $\mathcal{G}(H)$ iff $M \mathbf{p} = e_{\hat{\mathbf{1}}}$ and $\mathbf{p} \geq 0$, where the first $n$ entries of $e_{\hat{\mathbf{1}}}$ are $1$ and the rest $0$ (with $n$ denoting the number of rows of $M_{\mathrm{Normalisation}}$). 
Now, let $e_s$ be the vector with $|V(H)|$ entries, such that $e_s(\mathbf{x}) = 1$ if $\mathbf{x} \in s$ and $0$ otherwise, where recall that $s \subset V(H)$ is a subset of mutually-orthogonal events such that $\sum_{\mathbf{x} \in s} p(\mathbf{x}) = 1$ for all allowed behaviours in $\mathcal{G}(H)$. What we then need to show is that the vector $(e_s^\mathrm{T},1)$ is a linear combination of the rows of the augmented matrix $(M,e_{\hat{\mathbf{1}}})$. For this, consider a vector $\mathbf{r}$ in the kernel of $M$, such that $M \, \mathbf{r} = \vec{\mathbf{0}}$. Since $\mathbf{q}$ is a behaviour in $H$ that assigns non-zero values, then there exists a small enough $\epsilon$ such that both $\mathbf{p}_{+\epsilon} = \mathbf{q} + \epsilon \, \mathbf{r}$ is a valid behaviour, i.e., $M \, \mathbf{p}_{+\epsilon} = e_{\hat{\mathbf{1}}}$ and $\mathbf{p}_{+\epsilon} \geq 0$, and also $\mathbf{p}_{-\epsilon} = \mathbf{q} - \epsilon \, \mathbf{r}$ is a valid behaviour, i.e., $M \, \mathbf{p}_{-\epsilon} = e_{\hat{\mathbf{1}}}$ and $\mathbf{p}_{-\epsilon} \geq 0$. Now, this means that both $e_s^\mathrm{T} \mathbf{p}_{\pm \epsilon} =1$, from which follows that $2\epsilon e_s^\mathrm{T} \mathbf{r} = 0$. Since $\epsilon > 0$, then $e_s^\mathrm{T} \mathbf{r} = 0$ for any vector $\mathbf{r}$ in the kernel of $M$. The fact that $e_s$ lives in the  space orthogonal to the kernel of $M$ means that it is a linear combination of the rows of $M$. This, combined with the normalisation constraint $e_s^\mathrm{T} \mathbf{q} = 1$ proves that the rank condition (9b) of Ref.~\cite{sainz2018multipartite} is satisfied. Therefore, any subset $s \subset V(H)$ of mutually-orthogonal events such that $\sum_{\mathbf{x} \in s} p(\mathbf{x}) = 1$ is already a hyperedge in $H$, which concludes the proof that $\overbar{H} = H$. 
    
\end{proof}

\section{Algorithms}\label{ap:algo}

In this section we discuss how to in practice certify whether two contextuality scenarios admit of the same set of almost-quantum behaviours. For this we discuss how to verify that the conditions $\MN$ from Def.~\ref{def:MN} and $\AOL$ from Def.~\ref{def:aol} are satisfied. 

\begin{prop}\label{pr:sdpMN}
Consider two contextuality scenarios $H_1$ and $H_2$ with the same vertex set. Then whether their sets of almost-quantum behaviours satisfy $\MN$ can be checked with at most $2(|E(H_1)|+|E(H_2)|)$ semidefinite programs (SDPs).  

These SDPs are the following: 

\bigskip

For $e\in E(H_1)$ compute
\begin{align}
    f_{1,e} := \min_{\mathbf{p} \in \mathcal{Q}_1(H_2)} \left( \sum_{v \in e} p(v) \right)\,,\quad g_{1,e} := \max_{\mathbf{p} \in \mathcal{Q}_1(H_2)} \left( \sum_{v \in e} p(v) \right)\,.
\end{align}
For $e\in E(H_2)$ compute
\begin{align}
    f_{2,e} := \min_{\mathbf{p} \in \mathcal{Q}_1(H_1)} \left( \sum_{v \in e} p(v) \right)\,,\quad g_{2,e} := \max_{\mathbf{p} \in \mathcal{Q}_1(H_1)} \left( \sum_{v \in e} p(v) \right)\,.
\end{align}
If $f_{1,e}=1=g_{1,e}$ for all $e\in E(H_1)$ and $f_{2,e}=1=g_{2,e}$ for all $e\in E(H_2)$, then $\mathcal{Q}_1(H_1)$ and $\mathcal{Q}_1(H_2)$ satisfy $\MN$.
\end{prop}
It is worth mentioning that Prop.~\ref{pr:sdpMN} applies when both sets of almost-quantum behaviours are non-empty. When both sets are empty the question does not apply (the sets are trivially equal), and when only one is non-empty then the sets are not equal to begin with. 
\begin{proof}
Recall that the set of almost-quantum behaviours is defined via a collection of linear and positive-semidefinite constraints \cite[Def.~6.1.2]{acin2015combinatorial}, and hence optimising a linear function over it is an SDP. Therefore, computing the quantities $\{f_{1,e},g_{1,e}\}_{e \in E(H_1)}$ amounts to $2|E(H_1)|$ SDPs and computing the quantities $\{f_{2,e},g_{2,e}\}_{e \in E(H_2)}$ amounts to $2|E(H_2)|$. 

Now, if $f_{1,e}=1=g_{1,e}$ for all $e\in E(H_1)$ then this implies Eq.~\eqref{eq:MN2} is satisfied. Moreover, if $f_{2,e}=1=g_{2,e}$ for all $e\in E(H_2)$ then this implies Eq.~\eqref{eq:MN1} is satisfied. Hence, this algorithm based on a collection of SDPs tests the satisfaction of $\MN$ by almost-quantum behaviours. 
\end{proof}

\bigskip

It is worth mentioning that whether two contextuality scenarios admit the same sets of quantum behaviours is determined only by the $\MN$ property, as shown in Prop.~\ref{prop:necsufquan}. However, checking whether two sets of quantum behaviours satisfy $\MN$ is an undecidable problem (follows from Ref.~\cite{fritz2021quantum}), and hence no general algorithm to test it exists. 

\bigskip

For discussing $\AOL$ it is convenient to introduce the notion of the Theta Body of a graph \cite{GROTSCHEL1986330}, here denoted by $\TB$. 

\begin{defn}\label{def:TH} (Ref.~\cite{GROTSCHEL1986330}) \\
Let $G$ be a graph, and let $\mathbf{x}$ be an assignment of non-negative numbers to the vertices of $G$. The Theta Body of $G$ is defined as 
\begin{align}
    \TB(G) = \{\mathbf{x} \, | \, \vartheta(\overbar{G},\mathbf{x}) \leq 1\}\,,
\end{align}
where $\overbar{G}$ denotes the complement of $G$, and $ \vartheta$ is the weighted Lov\'asz number. 
\end{defn}
Notice that $\mathcal{Q}_1(H) = \mathcal{G}(H) \cap \TB(\mathrm{OrthoGraph}(H))$\,.

\begin{cor}
Consider two contextuality scenarios $H_1$ and $H_2$ with the same vertex set. Their sets of almost-quantum behaviours satisfy $\AOL$ if and only if 
\begin{align} \label{eq:incTB}
    \mathcal{Q}_1(H_1) \subseteq \TB(\mathrm{OrthoGraph}(H_2)) \quad \textrm{and} \quad \mathcal{Q}_1(H_2) \subseteq \TB(\mathrm{OrthoGraph}(H_1))\,.
\end{align}
\end{cor}
Checking Eq.~\eqref{eq:incTB} cannot generically be expressed as an SDP, although quantifier elimination methods might provide an exact (although less practical) finite algorithm for it. 

We see then that checking in general whether two contextuality scenarios admit of the same set of almost-quantum behaviours as per Prop.~\ref{prop:necsuf} is not generically easy. Notice, however, that when the two scenarios have the same orthogonality graphs then the problem simplifies substantially: $\AOL$ is immediately satisfied, and the algorithm of Prop.~\ref{pr:sdpMN} reduces to two SDPs as we see next. 

\begin{prop}\label{pr:sdpMNeo}
Consider two contextuality scenarios $H_1$ and $H_2$ with the same vertex set, and which satisfy $\mathrm{OrthoGraph}(H_1) = \mathrm{OrthoGraph}(H_2)$.  Then whether their sets of almost-quantum behaviours satisfy $\MN$ can be checked with two semidefinite programs (SDPs).  

These SDPs are the following: 

\bigskip

\begin{align}
    F_{1} &:= \max_{\mathbf{p} \in \mathcal{Q}_1(H_1)} \sum_{e \in E(H_2)} \left( 1 - \sum_{v\in e} p(v) \right)\,,\\
    F_{2} &:= \max_{\mathbf{p} \in \mathcal{Q}_1(H_2)} \sum_{e \in E(H_1)} \left( 1 - \sum_{v\in e} p(v) \right)\,.
\end{align}
If $F_1 = 0 = F_2$ then $\mathcal{Q}_1(H_1)$ and $\mathcal{Q}_1(H_2)$ satisfy $\MN$.
\end{prop}
\begin{proof}
Since $H_1$ and $H_2$ have the same orthogonality graphs, this means that the hyperedges of one are cliques in the orthogonality graph of the other. Hence, the Consistent Exclusivity principle \cite[Def.~7.1.1]{acin2015combinatorial} implies 
\begin{align}
\sum_{v \in e} p(v) &\leq 1 \,, \quad \forall \, \mathbf{p} \in \mathcal{Q}_1(H_1)\,,\, \forall \, e \in E(H_2) \,,\\
\sum_{v \in e} p(v) &\leq 1 \,, \quad \forall \, \mathbf{p} \in \mathcal{Q}_1(H_2)\,,\, \forall \, e \in E(H_1) \,.
\end{align}
Hence, we need to check that these sums can never be strictly lower than $1$. 

Notice that if $F_{1}=0$ this means that there is no hyperedge $e \in E(H_2)$ and no $\mathbf{p} \in \mathcal{Q}_1(H_1)$ for which $\sum_{v \in e} p(v) < 1$. Similarly, if $F_{2}=0$ this means that there is no hyperedge $e \in E(H_1)$ and no $\mathbf{p} \in \mathcal{Q}_1(H_2)$ for which $\sum_{v \in e} p(v) < 1$. Hence $F_1 = 0$ implies Eq.~\eqref{eq:MN1} holds, and $F_2 = 0$ implies Eq.~\eqref{eq:MN2} holds. Hence, $F_1 = 0 = F_2$ implies that $\mathcal{Q}_1(H_1)$ and $\mathcal{Q}_1(H_2)$ satisfy $\MN$.
Finally, notice that since the set of almost-quantum behaviours is defined via a collection of linear and positive-semidefinite constraints \cite[Def.~6.1.2]{acin2015combinatorial}, then computing the linear functionals $F_1$ and $F_2$ are each a single SDP. 

\end{proof}

\bigskip

Next we discuss an algorithm to check whether a collection of contextuality scenarios satisfy the hypothesis required in Lemma \ref{lem:sc}. This hypothesis is recalled in the definition below. 

\begin{defn}\label{def:proplem}
Consider a contextuality scenario $H$. We say that $H$ satisfies the hypothesis of Lemma \ref{lem:sc} if: for each pair of vertices $u,v \in V(H)$ such that $u \not\perp v$ (including $u=v$) there exists a behaviour $\mathbf{p}_{uv} \in \mathcal{G}(H)$ such that $p_{uv}(u)=1$ and $p_{uv}(v)>0$. 
\end{defn}

Checking whether $H$ satisfies the above mentioned hypothesis amounts to a single  simple linear program, which we state below. 

\begin{prop}\label{prop:LPlem}
Let $H$ be a contextuality scenario. $H$ satisfies the hypothesis of Lemma \ref{lem:sc}, articulated in Def.~\ref{def:proplem}, if and only if the following linear program is feasible and yields a strictly positive solution ($\delta^* > 0$): 

\begin{align}
    \text{maximise} & \quad \delta \\ \label{eq:normxu}
    \text{subject to} & \quad A \mathbf{x}_u = \mathbf{1} \quad \forall u \in V(H)\,,\\ \label{eq:nonnegxu}
    & \quad \mathbf{x}_u \geq 0 \quad \forall u \in V(H)\,,\\ \label{eq:u1xu}
    & \quad \mathbf{x}_u(u) = 1 \quad \forall u \in V(H)\,,\\ \label{eq:v0xu}
    & \quad \mathbf{x}_u(v) \geq \delta \quad \forall u \in V(H)\,, \, v \in N(u)\,,\\ \label{eq:deltam0}
    & \quad \delta \geq 0 \,,
\end{align}
where $\mathbf{x}_u$ is a vector in $\mathbb{R}^{|V(H)|}$, $N(u) = \{v \in V(H) \,|\, u \not\perp v\} \cup \{u\}$, $\mathbf{1}$ is a vector in $\mathbb{R}^{|E(H)|}$ whose entries are all $1$, and the matrix $A \in \mathbb{R}^{E(H)} \times \mathbb{R}^{V(H)}$ has rows indexed by the hyperedges of $H$ and columns indexed by the vertices of $H$ and its entries are $A_{ev} = 1$ if $v \in e$ and $0$ otherwise.

\end{prop}
\begin{proof}
The idea is that, for each $u\in V$, the vector $\mathbf{x}_u$ would result in a behaviour $\mathbf{p}_u$ that has $p_u(u) = 1$ and $p_u(v) > 0$ for all $v \in N(u)$, which is the desired property of Def.~\ref{def:proplem}. 

Let us begin with the `if' direction, and assume that the linear program is feasible and has a strictly positive solution. Eqs.~\eqref{eq:normxu} and \eqref{eq:nonnegxu} then imply that $\mathbf{x}_u \in \mathcal{G}(H)$ for all $u \in V$. Eq.~\eqref{eq:u1xu} then implies that $\mathbf{x}_u(u) = 1$ for all $u\in V$. Eq.~\eqref{eq:v0xu} and Eq.~\eqref{eq:deltam0} imply that $\mathbf{x}_u (v) > 0$ for all $v \in N(u)$. Hence, by identifying $\mathbf{p}_{uv} \equiv \mathbf{x}_u$ for all $v \in N(u)$ and for all $u \in V$ shows that $H$ satisfies Def.~\ref{def:proplem}. 

Now let us continue with the `only if' direction, and assume that for each $u \in V$ and each $v \in N(u)$ there exist behaviours $\mathbf{p}_{uv} \in \mathcal{G}(H)$ with the above-mentioned properties. Now define $\mathbf{x}_u = \frac{1}{|N(u)|}\sum_{v \in N(u)} \mathbf{p}_{uv}$. By convexity, $\mathbf{x}_u \in \mathcal{G}(H)$, and hence Eqs.~\eqref{eq:normxu} and \eqref{eq:nonnegxu} follow. Since $p_{uv}(u) = 1$ for all $v \in N(u)$ and for all $u \in V$, then Eq.~\eqref{eq:u1xu} holds. Now, since $p_{uv}(v) > 0$, this means that $\mathbf{x}_u(v) > 0$ for all $v \in N(u)$ and for all $u \in V$, and hence Eq.~\eqref{eq:v0xu} holds for a strictly positive $\delta$. Hence, Eq.~\eqref{eq:deltam0} is satisfied. All these imply that $\mathbf{x}_u$ defined as above yields a solution to the linear program with a strictly-positive $\delta$ given by 
\begin{align}
    \delta^* = \min_{u \in V\,,\, v \in N(u)} \mathbf{x}_u(v)\,.
\end{align}
\end{proof}

\section{Quasi-process functions and the disjunctive FR-product} \label{ap:hippo}

In this section we discuss how the disjunctive FR-product of general contextuality scenarios can give rise to functions that fail the logical consistency conditions for process functions \cite{kunjwal2023nonclassicality}, i.e., quasi-process functions. 

Ref.~\cite[Appendix G.3]{dourdent2025paradox} presents an ambiguous local product basis that generates a non-causal quasi-process function. For this, they take (i) a qubit system and measurements in the bases $\{\ket{0},\ket{1}\}$ and $\{\ket{+},\ket{-}\}$, and (ii) a qutrit system and measurements in the bases $\{\ket{0},\ket{1},\ket{2}\}$ and $\{\ket{+},\ket{-},\ket{2}\}$. For the qubit system, measurement in those basis are associated with the labels: $\ket{0} \, \rightarrow \, (0|0)$, $\ket{1} \, \rightarrow \, (1|0)$, $\ket{+} \, \rightarrow \, (0|1)$, $\ket{-} \, \rightarrow \, (1|1)$. For the case of the qutrit, the measurement events further include $\ket{2} \, \rightarrow \, (2|0)$ for the first measurement, and $\ket{2} \, \rightarrow \, (2|1)$ for the second. The fact that $\ket{2}$ appears in both of the qutrit's measurements makes the local basis `ambiguous', and bipartite states that are orthogonal may give rise to events that are not. The authors show that this gives rise to a quasi-process function $f$ defined by $x_1 = \delta_{a_2,2}$ and $x_2 = a_1$, where $x_1$ (resp.~$x_2$) and $a_1$ (resp.~$a_2$) denote the labels of the qubit's (resp.~   qutrit's) measurement and outcome. Indeed, $f$ is not a process function since it displays both the grandfather antinomy and the bootstrap antinomy \cite{dourdent2025paradox}.

Let us now see how to frame this example in the language of hypergraphs. On the one hand (i.e., the qubit) we have two dichotomic measurements with no overlapping outcomes -- we denote this by $H_A$. On the other hand (i.e., the qutrit), we have two three-outcome measurements that share one outcome -- we denote this by $H_B$. Here we denote their vertices and hyperedges as: 
\begin{align}
    V(H_A) &= \{0,1,+,-\}\,,\, E(H_A) = \{\{0,1\},\{+,-\}\}\,, \\
    V(H_B) &= \{0,1,+,-,2\}\,,\, E(H_B) = \{\{0,1,2\},\{+,-,2\}\}\,.
\end{align}

We will now argue that the set of vertices $S \subseteq V(H_A) \times V(H_B)$ defined by 
\begin{align}
    S = \{(0,0),(0,1),(1,+),(1,-),(+,2),(-,2)\}
\end{align}
is a hyperedge in $H_A \, {}^\disj \otimes H_B$ and gives rise to the quasi-process function $f$. 
On the one hand, direct inspection tells us that $S$ is composed of mutually orthogonal events. On the other hand, normalisation and non-signalling constraints imply that for any behaviour $\mathbf{p} \in \mathcal{G}(H_A \, {}^\disj \otimes H_B)$ the following holds:
\begin{align}
    \sum_{v\in S} p(v) &= p(0,0)+p(0,1)+p(1,+)+p(1,-)+p(+,2)+p(-,2) \,,\\
    &= [p_A(0) - p(0,2)] + [p_A(1) - p(1,2)] + p_B(2) \,,\\
    &= [p_A(0) - p(0,2)] + [p_A(1) - p(1,2)] + [p(0,2)+p(1,2)] \, \\
    &= p_A(0) + p_A(1) =  1\,.
\end{align}
Hence, $\sum_{v\in S} p(v) = 1$ for all $\mathbf{p} \in \mathcal{G}(H_A \, {}^\disj \otimes H_B)$. In addition, by explicitly constructing the matrices $M_{\mathrm{Normalisation}}$ and $M_{\mathrm{NoSignalling}}$, one can check that the rank condition of Ref.~\cite{sainz2018multipartite} is satisfied. Altogether, these imply $S \in E(H_A \, {}^\disj \otimes H_B)$. Also notice that $S \not\in E(H_A \, {}^\mathrm{max} \otimes H_B)$. This follows from the fact that neither party has a single fixed measurement choice in $S$. 

Let us now see how to understand $S$ as a quasi-process function. For this, let us first map the labels of $V(H_A \, {}^\disj \otimes H_B)$ to labels of $V(\mathcal{B}_{1,2,2}) \times V(\mathcal{B}_{1,2,3})$\,. Leveraging the ambiguity that outcome $2$ can occur in two different measurements of $H_B$, we exploit this freedom to assign the following: 
\begin{center}
    \begin{tabular}{c|c}
    Events in $S$ & Events in $V(\mathcal{B}_{1,2,2}) \times V(\mathcal{B}_{1,2,3})$ \\ \hline
    $(0,0)$ & $(00|00)$ \\
     $(0,1)$ & $(01|00) $\\
     $(1,+)$ & $(10|01)$\\
     $(1,-)$ & $(11|01)$\\
     $(+,2)$ & $(02|10)$ \\
     $(-,2)$ & $(12|11)$ 
\end{tabular}
\end{center}
Notice that each possible pair of outcomes $\{0,1\} \times \{0,1,2\}$ occurs only once in the above list of Bell events. One can then define a quasi-process function by noting that these events satisfy (i) $x_A = 0$ if $a_B \in \{0,1\}$ and $x_A = 1$ otherwise, and (ii) $x_B = a_A$. Hence, the quasi-process function it defines corresponds exactly with $f$ above. 
Finally, notice that $S$ does not correspond to any hyperedge in $ H_A \, {}^\textrm{max} \otimes H_B$, nor to the image of any of the hyperedges of $\mathcal{B}_{1,2,2} \, {}^\textrm{max} \otimes \mathcal{B}_{1,2,3}$ 
after identifying $(2|0) \leftrightarrow (2|1)$. 

We see then that hyperedges in the disjunctive product of two hypergraphs that are not local Bell scenarios can already give rise to hyperedges which may be associated to functions that are not process functions but rather quasi-process functions. 

\bigskip

It is worth mentioning that a similar conclusion can be reached from a simpler example: 
\begin{align}
    V(H) = \{a,b,c\}\,,\, E(H) = \{\{a,b\},\{a,c\}\}\,,
\end{align}
and noting that the set $S = \{(a,a),(a,b),(b,a),(c,c)\}$ is (i) a hyperedge in $H {}^\disj \otimes H$, (ii) is not a hyperedge in $H {}^\mathrm{max} \otimes H$, and (iii) defines a quasi-process function $(x_1,x_2):=(a_2,a_1)$ that features both the grandfather antinomy and the bootstrap antinomy. The proof technique for these claims is analogous to those used above in this section:
\begin{compactitem}
\item To see that $S$ is a hyperedge $H {}^\disj \otimes H$, notice that (i) it is composed of pairwise-orthogonal events, (ii) $\sum_{v \in S} p(v) = 1$ for any behaviour $\mathbf{p} \in \mathcal{G}(H {}^\disj \otimes H)$ since no-signalling implies that $p(b,b)=p(c,c)$,  and (iii) the rank condition from Ref.~\cite{sainz2018multipartite} is also satisfied, as can be checked by constructing the corresponding normalisation and no-signalling matrices (like in the proof of Lemma \ref{lem:sc}). 
\item To see that $S$ is not a hyperedge in $H {}^\mathrm{max} \otimes H$, notice that no party in $S$ has a single measurement choice (necessary condition for the first party in a measurement protocol). 
\item $S$ defines a quasi-process function $(x_1,x_2):=(a_2,a_1)$, since its events can be mapped to events in $V(\mathcal{B}_{1,2,2}) \times V(\mathcal{B}_{1,2,2})$ as: $(a,a) \rightarrow (00|00)$, $(a,b) \rightarrow (01|10)$, $(b,a) \rightarrow (10|01)$, $(c,c) \rightarrow (11|11)$\,.
\end{compactitem}

\end{document}